\documentclass[pdflatex,sn-nature]{sn-jnl}

 \usepackage{placeins}
\usepackage{rotating}
\usepackage{anyfontsize}
\usepackage{graphicx}%
\usepackage{multirow}%
\usepackage{amsmath,amssymb,amsfonts}%
\usepackage{amsthm}%
\usepackage{mathrsfs}%
\usepackage[title]{appendix}%
\usepackage{xcolor}%
\usepackage{textcomp}%
\usepackage{manyfoot}%
\usepackage{booktabs}%
\usepackage{algorithm}%
\usepackage{algorithmicx}%
\usepackage{algpseudocode}%
\usepackage{listings}%
\usepackage{lineno}
\usepackage{gensymb}
\usepackage{caption}

\begin{document}

\title[Article Title]{Molecular diversity as a biosignature}

\author*[1]{\fnm{Gideon} \sur{Yoffe}}\email{gidi.yoffe@weizmann.ac.il}

\author[2]{\fnm{Fabian} \sur{Klenner}}\email{fabiank@ucr.edu}

\author[3]{\fnm{Barak} \sur{Sober}}\email{barak.sober@mail.huji.ac.il}

\author[1]{\fnm{Yohai} \sur{Kaspi}}\email{yohai.kaspi@weizmann.ac.il}

\author[1]{\fnm{Itay} \sur{Halevy}}\email{itay.halevy@weizmann.ac.il}

\affil[1]{\orgdiv{Department of Earth and Planetary Sciences}, \orgname{Weizmann Institute of Science}, \orgaddress{\city{Rehovot}, \postcode{76100}, \country{Israel}}}

\affil[2]{\orgdiv{Department of Earth and Planetary Sciences}, \orgname{University of California, Riverside}, \orgaddress{\city{Riverside}, \state{CA}, \postcode{92521}, \country{USA}}}

\affil[3]{\orgdiv{Department of Statistics and Data Science}, \orgname{The Hebrew University of Jerusalem}, \orgaddress{\city{Jerusalem}, \postcode{9190501}, \country{Israel}}}


\abstract{
The search for life in the Solar System hinges on data from planetary missions. Detecting biosignatures based on molecular identity, isotopic composition, or chiral excess requires measurements that current and planned missions can only partially provide.
We introduce a new class of biosignatures, defined by the statistical organization of molecular assemblages and quantified using diversity metrics. Using this framework, we analyze amino-acid diversity across a dataset spanning terrestrial and extraterrestrial contexts.
We find that biotic samples are consistently more diverse---and therefore distinct---than their sparser abiotic counterparts. This distinction also holds for fatty acids, indicating that the diversity signal reflects a fundamental biosynthetic signature. It also proves persistent under modeled space-like degradation.
Relying only on relative abundances, this biogenicity assessment strategy is applicable to any molecular composition data from archived, current, and planned planetary missions. By capturing a fundamental statistical property of life’s chemical organization, it may also transcend biosignatures that are contingent on Earth’s evolutionary history.
}

\maketitle

Life as we know it is built from a finite repertoire of organic molecules. Among these, amino acids, lipids, and related amphiphiles (molecules containing both water-attracting and water-repelling parts) occupy a privileged position, as the former are the building blocks of proteins and the latter are essential for membrane structure and function.\cite{deamer1986role} The compositions and abundance patterns of such molecular classes are therefore widely regarded as key targets in the search for life beyond Earth.\cite{klenner2020discriminating} Yet, these compounds are not exclusive to biology: they have been detected in meteorites and asteroids,\cite{martins2007indigenous,burton2012propensity, burton2013extraterrestrial, burton2014effects, parker2023extraterrestrial, Glavin2025} simulated prebiotic environments,\cite{kebukawa2017one} and terrestrial settings where abiotic synthesis cannot be ruled out.\cite{klevenz2010concentrations}

The discovery that non-biological processes generate diverse organic mixtures under varied conditions has informed multiple origin-of-life hypotheses. These include endogenous emergence in geochemically active settings, such as hydrothermal systems,\cite{marshall1994hydrothermal} impact-heated basins,\cite{cockell2006origin} transient ice--water interfaces,\cite{kanavarioti2001eutectic} and carbonate lakes undergoing wet--dry cycles.\cite{toner2020carbonate} Other conjectures support exogenous delivery from space, consistent with the widespread occurrence of organic molecules in extraterrestrial bodies.\cite{deamer2002first}
Although organic compounds form in both abiotic and biotic contexts, the properties they acquire are governed by profoundly different constraints. Abiotic synthesis is shaped primarily by thermodynamics and reaction kinetics,\cite{higgs2009thermodynamic} whereas biological synthesis is organized by metabolism and functional demand.\cite{smith2004universality} These differences make organic assemblages a natural substrate for biosignature inference.

Among organic biosignatures, chirality is the most iconic. Life on Earth synthesizes almost exclusively L-enantiomers (the left-handed form of chiral molecules), whereas abiotic pathways yield racemic mixtures (unbiased distributions of right- and left-handed forms), so an L-excess is often treated as a biosignature.\citep{patty2018remote} Yet this signal is fragile. Detecting chiral asymmetry at low concentrations requires precise, contamination-free protocols that are often infeasible in situ, and racemization erodes the asymmetry over time.\cite{schopf1968amino} Isotopic enrichment is another classical biosignature with similar limitations. Biological processes can preferentially incorporate one isotope over another, producing characteristic isotope ratios in organic matter, but these ratios can be obscured by abiotic exchange and diagenesis and are likewise difficult to measure in situ with sufficient precision.\cite{sephton2002organic}

Beyond specific molecular and isotopic markers, origin-diagnostic information also resides in the statistical organization of organic assemblages. For amino acids, abiotic synthesis tends to favor low-mass, structurally simple compounds, such as glycine and alanine,\cite{kebukawa2017one} because each added carbon or functional group typically incurs an energetic cost, reducing the relative abundance of more complex species.\cite{higgs2009thermodynamic} Biosynthesis, by contrast, can bypass this hierarchy: enzymatic control enables the targeted production of complex species in proportions determined by physiological function.\cite{akashi2002metabolic} This decoupling between complexity and abundance reflects a defining feature of life: sustained energetic investment to maintain chemical distributions not favored at equilibrium.\cite{england2013statistical} The maintenance of complex molecular distributions thus appears to be a more fundamental property of biological systems.
An analogous principle operates in fatty acids, where biology selects a narrow subset of chain lengths for membrane function, while abiotic synthesis yields chain-length distributions agnostic to such functional constraints.\cite{ohlrogge1997regulation, mccollom1999lipid}

This view is supported by prior empirical work showing that amino-acid and lipid assemblages encode diagnostically meaningful structure beyond individual compounds, both in meteoritic samples affected by diagenesis and in laboratory-scale analytical discrimination, thereby motivating distribution-level biosignature metrics.\citep{buckner2024quantifying} Experimental studies further show that microbial metabolism imposes characteristic distributional fingerprints, most notably through selective glycine depletion.\citep{schwendner2022microbial} Agnostic biosignatures based on collective molecular patterns provide a complementary strategy, but often require calibration to specific instruments and measurement conditions.\citep{cleaves2023robust}
However, such attribution may be ambiguous due to degradation and alteration. On Earth and Mars, amino-acid profiles may be altered by contamination, oxidation, or thermal degradation.\cite{schopf1968amino} Even in anoxic subsurface settings, selective loss continues, particularly at elevated temperature and pressure.\cite{klevenz2010concentrations}
In space, ultraviolet radiation and energetic plasma drive photolysis and radiolysis, selectively degrading molecules based on their structure.\cite{yoffe2025fluorescent} These processes are especially active on unshielded surfaces. 
Extant organic assemblages, therefore, reflect both synthetic origin and accumulated environmental processing.

\begin{figure*}
\centering
\rotatebox[origin=c]{0}{\includegraphics[scale = 1.]{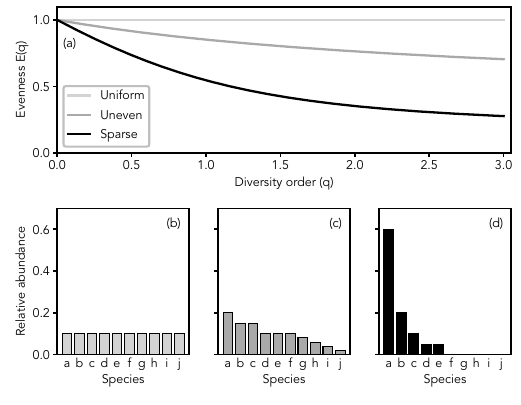}}
\caption{\textbf{Illustrative evenness curves and abundance profiles.}
(\textbf{a}) Evenness curves for three assemblages of ten species distributed uniformly (light gray), unevenly (dark gray), and sparsely (black), computed across diversity orders ($q$). Flatter curves indicate more evenly distributed species.  
(\textbf{b}–\textbf{d}) Corresponding species-abundance distributions for the three assemblages are shown with the same color scheme.}
\label{fig: evenness_richness_illustration}
\end{figure*}

To move beyond reliance on specific molecular identities, stereochemical signals, or instrument-specific limitations, we introduce a statistical framework for analyzing organic assemblages based on the \textit{ecodiversity} formalism.\cite{chao2014unifying} Ecodiversity statistics, originating in ecological theory, quantify the structure of biological communities and have hitherto not been applied to molecular inventories. These measures capture the number of unique species and the distribution of their abundances. 
We treat organic assemblages collected within a unified context (e.g., from the same asteroid) as individual samples. Within a given molecular family, the set of detected species in a sample, together with their relative abundances, defines an assemblage. The distribution of detected species within each sample is therefore the primary data object. We characterize this statistical structure using two metrics: \textit{richness}, which is the number of distinct species in the sample, and \textit{diversity}, which measures the uniformity of their relative abundances. A sample with many species but dominated by a few has high richness but low diversity. These metrics are jointly expressed using Hill numbers---a parametric family of diversity indices (see Methods).\citep{hill1973diversity} Varying sensitivity to rare versus dominant species yields a continuous diversity profile. Normalizing each profile by its sample-wise richness defines the evenness curve. This normalization makes evenness invariant to species count and reflects only the structure of relative abundances rather than absolute diversity. Evenness curves are therefore agnostic to species identity and measurement units, enabling direct comparison between samples with disparate richness or total concentration. Each assemblage is thus represented as a value vector encoding the distribution’s shape but not its magnitude, isolating compositional structure from inventory size (Fig.~\ref{fig: evenness_richness_illustration}).

Each amino-acid abundance is accompanied by an uncertainty estimate reflecting variability introduced during extraction and quantification. Where reported, uncertainties are used directly; otherwise, they are inferred from variation among compositionally similar profiles within the same study. This represents each sample as an ensemble of plausible distributions from which evenness-curve distributions are generated, propagating measurement uncertainty through the diversity framework without imposing uniform assumptions across datasets.
Dissimilarities between samples are then computed as pairwise distances between their evenness-curve distributions, expressed as $z$-scores—the likelihood that two samples originate from the same distribution in units of standard deviation under a normal model. The procedures for computing evenness curves, propagating uncertainty, and quantifying pairwise dissimilarities are described in the Methods.

\section*{Mapping sample dissimilarities across origins} \label{sec: dissimilarity_with_glycine}

\subsection*{Amino acids}

We apply the framework to a curated, deliberately heterogeneous dataset of amino-acid assemblages spanning terrestrial, extraterrestrial, and experimental contexts. Biotic assemblages comprise environmental amino-acid distributions shaped by biological signatures in microbial cultures, marine and estuarine sediments, and fossilized biota. Abiotic assemblages include meteoritic and asteroidal materials, simulated icy-moon analog profiles, and laboratory-synthesized materials reflecting early Solar System and prebiotic chemistry. Across categories, samples encode distinct combinations of source chemistry, alteration history, preservation regime, and analytical extraction methodology. Sample origin and context are summarized in Supplementary Table~1, pairwise dissimilarities are shown in Extended Data Fig.~1, and the per-sample uncertainty models are listed in the Supplementary Information.

Biotic samples are labeled by the dominant biological source: MIC denotes microbially derived distributions, EUK denotes eukaryotic material, and MIX denotes mixed-source environmental organics, where contributions cannot be attributed to a single source (for example, sediments containing both microbial and detrital inputs). Abiotic samples are labeled by provenance: meteorites and returned asteroidal materials use standard carbonaceous chondrite group tags, and U denotes ureilite-hosted assemblages. When profiles were measured using multiple extraction techniques, we considered the total hydrolyzable amino acids (THAA): all amino acids released through hydrolysis, whether originally free or polymer-bound. This provides a more complete inventory than free amino-acid measurements alone. The compiled dataset is intended to capture robust distribution-level patterns rather than artifacts of specific environments, ages, or analytical protocols. Samples are grouped into three categories: biotic, abiotic, and mixed. The biotic category includes modern and ancient terrestrial assemblages with confirmed or inferred biological origin. The abiotic category comprises synthetic, simulated, and uncontaminated extraterrestrial samples with no evidence of biological input. The mixed category includes meteorites with possible terrestrial contamination and terrestrial samples where both biotic and abiotic contributions are plausible.

\begin{figure*}[b!]
\centering
\rotatebox[origin=c]{0}{\includegraphics[scale = 0.8]{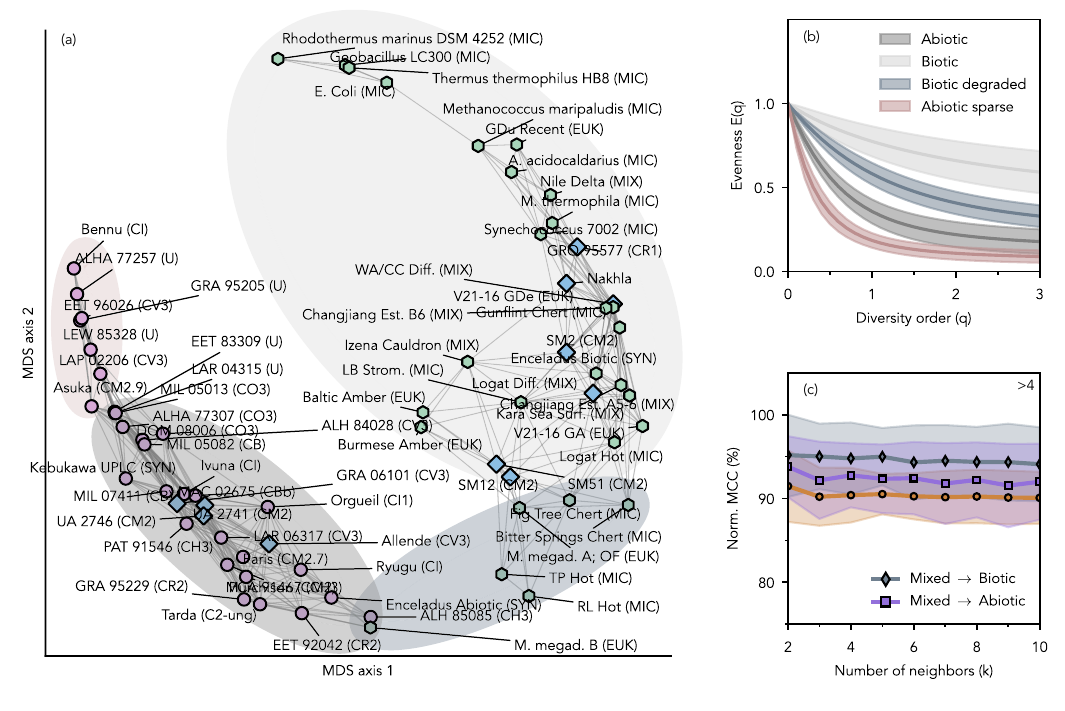}}
\caption{\textbf{Dissimilarity analysis of evenness curves for amino-acid assemblages.}
(\textbf{a}) Multidimensional Scaling (MDS) projection of pairwise dissimilarities between evenness curves, $E(q)$. Each point represents a sample; distances increase with statistical separation. Edges connect samples to the 25$^{\mathrm{th}}$ percentile of their nearest neighbors. Markers denote inferred origin: biotic (green hexagons), abiotic (pink circles), and mixed (blue diamonds).  
(\textbf{b}) Evenness-curve distributions for four sample groups. Solid lines indicate group means; shaded regions denote one standard deviation. Each color represents the distribution of samples contained within the shaded regions of the same color in panel (a).  
(\textbf{c}) Classification performance of sample origin using $k$-Nearest-Neighbors ($k$NN) applied to pairwise dissimilarities projected onto the first two MDS axes. Three labeling schemes are evaluated: all three groups retained, and two alternatives in which the mixed group is assigned to either the biotic or abiotic class. Accuracy is reported as the normalized Matthews Correlation Coefficient (MCC), where 50\% corresponds to random assignment and 100\% to perfect classification. Uncertainty is estimated by bootstrapping class-balanced subsamples and recomputing $k$NN accuracy for each subsample. The value shown at the top right denotes the smallest permutation-based $z$-score across all $k$. It measures the deviation of the observed mean MCC from the mean under permuted class labels, using the most conservative value across the three labeling schemes.}
\label{fig: results_aa}
\end{figure*}

Two self-consistent clusters emerge (Fig.~\ref{fig: results_aa}a). The first contains predominantly biotic samples: fresh biomass extracts,\citep{cordova201713c, beck2018measuring, simensen2022experimental, jiang2024, wang2024microbial} modern and ancient sediments from marine, depositional, and hydrothermal settings,\cite{king1972amino, klevenz2010concentrations, fuchida2015concentrations, wei2022variability, gaye2022aaah} Precambrian fossil-bearing rocks,\cite{schopf1968amino} Jurassic stromatolites,\cite{ballarini1994amino} and Cretaceous fossils.\cite{mccoy2019ancient, saitta2024non} The second cluster contains explicitly abiotic samples, including carbonaceous chondrites spanning a range of parent-body processing states,\citep{martins2007indigenous, burton2012propensity, burton2013extraterrestrial, burton2014effects, glavin2020asuka, Glavin2025} material returned from the aqueously altered asteroid Ryugu,\citep{parker2023extraterrestrial} a prebiotic synthesis experiment (Kebukawa~UPLC),\cite{kebukawa2017one} and a synthetic Enceladus-ocean analogue.\citep{klenner2020analog} Within this group, ALH~85085 forms the most diverse end-member, reflecting a CH3 parent-body history that allows multiple amino-acid formation pathways while limiting subsequent aqueous or thermal homogenization.\citep{burton2013extraterrestrial}

Samples labeled “mixed” fall clearly into either cluster. UA~2741 and UA~2746, fragments of the Aguas~Zarcas meteorite,\cite{glavin2021az} are reported as potentially contaminated yet cluster with abiotic samples. By contrast, samples from GRO~95577\cite{martins2007indigenous} and Nakhla\cite{glavin1999nakhla}---both heavily contaminated---cluster with the biotic group. The same holds for diffuse fluids from the Wideawake and Comfortless Cove (WA/CC~Diff.) and Logatchev (Logat~Diff.) hydrothermal fields.\cite{klevenz2010concentrations} In these cases, the biotic signal appears to dominate. The Allende meteorite sample groups with abiotic samples but lies closer to the biotic cluster, consistent with reported potential terrestrial contamination.\citep{burton2012propensity} The Sutter's Mill fragments SM2, SM12, and SM51 were likewise reported to have experienced terrestrial contamination,\citep{burton2014effects} and cluster with the biotic group.

Within the abiotic cluster, we identify a sparse sub-group (“Abiotic sparse”) comprising minimally aqueously altered type-3 carbonaceous chondrites, ungrouped carbonaceous meteorites, ureilites, and material recently analyzed from asteroid Bennu.\citep{burton2012propensity, Glavin2025} These samples represent environments dominated by limited aqueous synthesis or high-temperature processing, which restrict the production and retention of diverse soluble amino acids. Their inventories are therefore characterized by a strong dominance of a few simple species. This gradient suggests a continuum in abiotic amino-acid inventories driven primarily by parent--body processing, with terrestrial overprinting superposed: extensive aqueous alteration and low-temperature synthesis tend to produce richer, more even mixtures, whereas minimal alteration or high-temperature processing yields sparse distributions. This behavior is evident even among CI carbonaceous samples that share similar bulk mineralogy yet exhibit markedly different diversity signals.\citep{Glavin2025}

Between the clusters lies a buffer of samples with ambiguous diversity signatures (“Biotic degraded”). These include TP~Hot, RL~Hot, and Logat~Hot, fluids from high-temperature hydrothermal vents ($\gtrsim$350$^\circ$C) at the Turtle~Pits, Red~Lion, and Logatchev fields, where organic material derives from microbial ecosystems and thermally altered biological debris.\cite{klevenz2010concentrations} A similar ambiguous pattern appears in Precambrian fossil-bearing cherts from the Fig~Tree Group ($\gtrsim$3.1 Ga) and Bitter~Springs Formation ($\sim$1 Ga),\cite{schopf1968amino} and in two samples of \textit{Megaloolithus megadermus} dinosaur eggshell---M.~megad.~B and M.~megad.~A;~OF---from the Late Cretaceous ($\sim$70 Ma).\cite{saitta2024non} The former sample preserves more endogenous signal, whereas the latter, derived from surface flakes, appears depleted. The Bitter~Springs chert likewise retains a clearer microbial signature, whereas the older Fig~Tree chert appears more degraded. These differences suggest diagenetic loss and post-depositional overprinting.

That this group spans fossil-bearing rocks,\cite{schopf1968amino} fossilized biominerals,\citep{saitta2024non} and modern hydrothermal fluids\cite{klevenz2010concentrations} indicates that selective molecular loss can shift biotic samples toward sparse, abiotic-like diversity signatures across diverse environments. Nevertheless, evenness-curve distributions of the four clusters are consistently separated (Fig.~\ref{fig: results_aa}b): biotic samples are more uniform, abiotic ones markedly sparse, and the biotic-degraded group intermediate. Despite the diversity of sample contexts, classification remains robust. A $k$-nearest-neighbors analysis (Fig.~\ref{fig: results_aa}c) yields classification accuracies of $\sim$90–100.

\subsection*{Fatty acids}

We extend the diversity framework to a smaller dataset of biotic and abiotic fatty-acid assemblages. Abiotic samples were restricted to chain lengths greater than six carbon atoms (C$_{>6}$), as fatty acids in this range can self-assemble into membrane-like structures and are therefore comparable to biotic lipid inventories.\citep{deamer1986role} Sample origin and context are summarized in Supplementary Table~2, pairwise dissimilarities are shown in Extended Data Fig.~\ref{fig: results_aa}, and the corresponding uncertainty models are described in the Supplementary Information.

\begin{figure*}[b!]
\centering
\rotatebox[origin=c]{0}{\includegraphics[scale = 0.8]{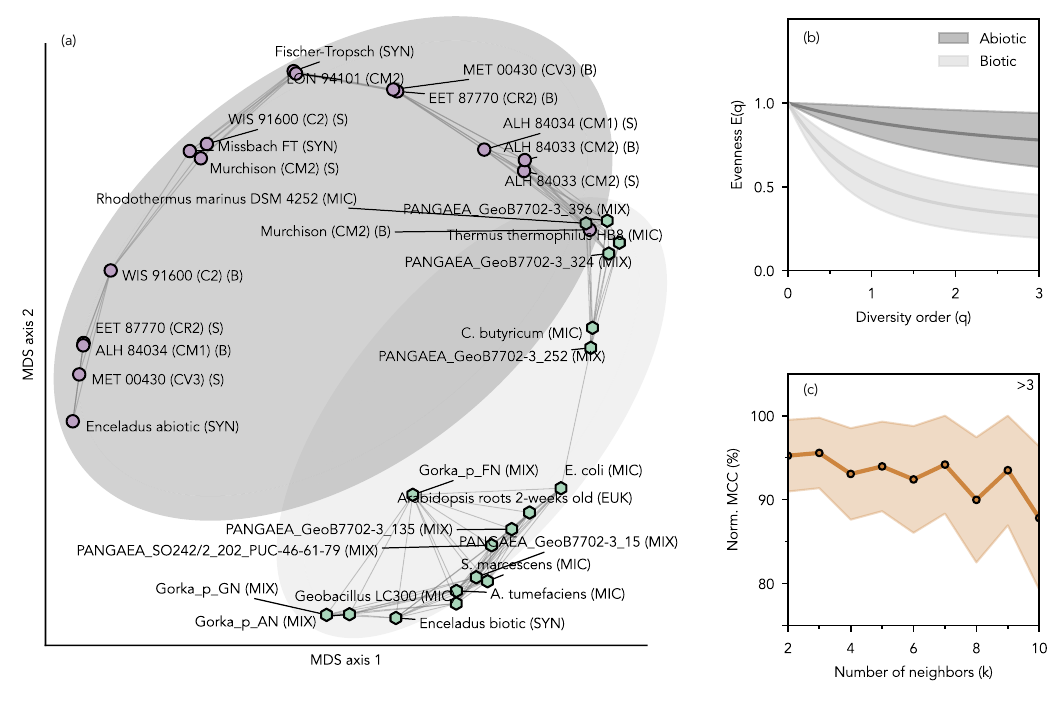}}
\caption{\textbf{Dissimilarity analysis of evenness curves for fatty-acid assemblages.}
(\textbf{a}) Multidimensional Scaling (MDS) projection of pairwise dissimilarities between evenness curves, analogous to the amino-acid analysis in Fig.2. Some meteoritic samples include suffixes indicating the fatty-acid subset used in the analysis: (S) and (B) denote straight- and branched-isomer subsets, respectively.  
(\textbf{b}) Evenness-curve distributions for four sample groups. Solid lines indicate group means; shaded regions denote one standard deviation.  
(\textbf{c}) Classification performance of sample origin using $k$-Nearest-Neighbors ($k$NN) applied to pairwise dissimilarities projected onto the first two MDS axes, analogous to the amino-acid analysis in Fig.~\ref{fig: results_aa}. Uncertainty is estimated by bootstrapping class-balanced subsamples and recomputing $k$NN accuracy for each subsample. The value shown at the top right denotes the smallest permutation-based $z$-score across all $k$. It measures the deviation of the observed mean MCC from the mean under randomized class labels.}
\label{fig: results_fa}
\end{figure*}

In this domain as well, biotic and abiotic profiles occupy distinct regimes with high statistical fidelity (Fig.~\ref{fig: results_fa}a). The diversity contrast reverses relative to amino acids: abiotic fatty-acid assemblages are more even across chain lengths (Fig.~\ref{fig: results_fa}b), consistent with broad production pathways such as Fischer–Tropsch synthesis.\cite{ohlrogge1997regulation} Biotic samples are sparser, reflecting membrane biosynthesis, which selects a limited subset of chain lengths and parities required for cellular function.\cite{mccollom1999lipid} Because amino acids and fatty acids reflect distinct but coupled biochemical requirements, separation across both classes points to a shared distribution-level imprint of biological coordination rather than a compound-class-specific artifact.

Meteoritic fatty acids further resolve into two statistically distinct populations: straight-chain (C$_{7–10}$) and branched (C$_{6–9}$) isomers. When analyzed separately, each subset forms a coherent evenness pattern and remains distinct from biotic samples (Fig.~\ref{fig: results_fa}c). When pooled into a single family, however, the diversity signal becomes dominated by differences in total abundance between the subsets. This behavior is consistent with meteoritic studies showing that straight and branched acids need not share a single production–processing history, and that branched-to-straight ratios vary with chain length and alteration state.\cite{lai2019meteoritic} Isotopic evidence likewise indicates systematically different fractionation signatures for the two moieties,\cite{alexander2017nature} supporting partially decoupled formation and alteration pathways. Further details of the fatty-acid diversity analysis are provided in the Supplementary Information.

\section*{Resolving biotic sample histories} \label{sec: biotic_sample_histories}

\begin{figure*}
\centering
\rotatebox[origin=c]{0}{\includegraphics[scale = 0.8]{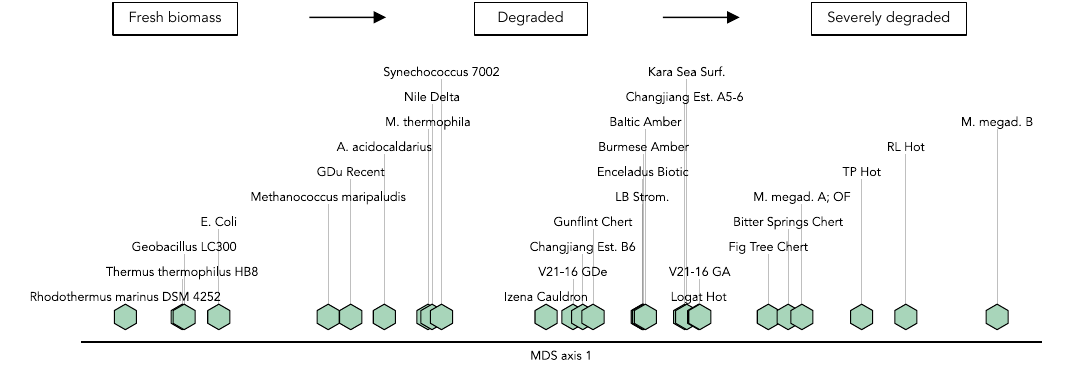}}
\caption{\textbf{One-dimensional projection of amino-acid samples of biotic origins dissimilarities.}
One-dimensional MDS projection of dissimilarities between evenness curves of samples of \textbf{biotic origin}, marked by green hexagons as in Figs.~2 and 3. Positions along the axis reflect relative compositional similarity in diversity structure. Samples span a continuum from fresh biomass with high diversity to progressively degraded assemblages with lower diversity. The axis does not represent a calibrated degradation scale, and no discrete thresholds are defined for preservation states.}
     \label{fig: result_1D_mds_biotic}
\end{figure*}

Fig.~\ref{fig: result_1D_mds_biotic} shows amino-acid samples of inferred biotic and mixed origins arranged along a gradient that broadly reflects compositional preservation. At one end lie fresh microbial biomass extracts with the highest diversity (Fig.~\ref{fig: results_aa}b), followed by recently deposited, well-preserved assemblages such as estuarine sediments and modern marine microfossils, where organic input is comparatively recent and has undergone little diagenesis. At the opposite end are samples subjected to extensive alteration in three distinct settings. These include fossil calcitic biominerals preserving organic matter over millions of years,\cite{saitta2024non} high-temperature hydrothermal fluids with intense water–rock interaction,\cite{klevenz2010concentrations} and ancient sediments that experienced prolonged diagenesis and recrystallization.\cite{schopf1968amino} Despite their differing contexts, all converge toward low internal diversity, consistent with progressive molecular loss during degradation.

Between these extremes lie samples from older marine sediments,\cite{klevenz2010concentrations, fuchida2015concentrations, wei2022variability} pelagic stromatolites in condensed carbonate sequences,\cite{ballarini1994amino} hydrothermal sediments with moderate thermal exposure,\cite{klevenz2010concentrations} and fossil inclusions preserved in amber.\cite{mccoy2019ancient} These assemblages retain a partial biotic imprint, though weaker than in better-preserved cases. Some positions along this axis reflect intrinsic compositional differences rather than degradation alone. For example, V21~16~GA (\textit{Globoquadrina altispira}) and V21~16~GDe (\textit{Globoquadrina dehiscens}), two foraminifera from the same Miocene horizon ($\sim$18 Ma), differ markedly: GA clusters with degraded samples, whereas GDe groups with more pristine assemblages.\cite{king1972amino} This distinction reflects differences in intrinsic species-level diversity in addition to degradation effects.\cite{king1972amino}

A similar pattern appears among high-temperature hydrothermal fluids. TP~Hot and RL~Hot exhibit sparse profiles, whereas Logat~Hot remains comparatively diverse.\cite{klevenz2010concentrations} This contrast may reflect environmental differences: Logatchev fluids circulate through serpentinized ultramafic rocks that can stabilize organic compounds, allowing richer inventories to persist despite similar temperatures.\cite{klevenz2010concentrations} A third example is the Gunflint~Chert sample, which, despite its age ($\sim$1.9 Ga), clusters with better-preserved assemblages.\cite{schopf1968amino} In contrast, the older Fig~Tree and younger Bitter~Springs cherts lie closer to degraded samples, likely reflecting differences in original microfossil communities and diagenetic history.\cite{schopf1968amino} Together, these cases illustrate the sensitivity of the diversity signal to variation within the biotic group. Differences in original composition, depositional setting, and post-depositional history all contribute to the observed structure.

\subsection*{A signature of biological organization} \label{sec: discussion}

We identify a persistent statistical distinction between biotic and abiotic amino-acid assemblages (Fig.~\ref{fig: results_aa}). Across terrestrial, extraterrestrial, and experimental datasets, biological samples exhibit greater internal evenness than abiotic counterparts. Abiotic assemblages are typically sparse and dominated by low-mass species, consistent with thermodynamic and kinetic constraints. This separation persists even when molecular identities are ignored, indicating that the diversity signal captures a structural property of biological organization. Despite heterogeneity and partial degradation, biotic and abiotic samples remain separable with high statistical confidence. This contrast reflects a defining feature of living systems: the coordinated synthesis and regulation of chemical diversity. Biosynthetic networks generate broad repertoires of molecules in functionally tuned proportions as a consequence of evolved metabolic organization.\cite{smith2004universality} Abiotic synthesis instead follows thermodynamic or kinetic optima, favoring selective production of small, stable compounds in narrower distributions.\cite{kebukawa2017one} The distinction thus lies not only in which molecules are present, but in how their abundances are organized.

The same framework reveals a complementary pattern in fatty acids (Fig.~\ref{fig: results_fa}). Here, the biotic–abiotic contrast reverses: biological samples are less even, whereas abiotic samples are more uniform across chain length. This follows from biological function. Proteins require a broad amino-acid repertoire, whereas membranes rely on a narrower subset of fatty acids defined by chain lengths used by cellular membranes.\cite{deamer1986role} Life thus expands diversity in one molecular class while constraining it in another.\cite{ohlrogge1997regulation, mccollom1999lipid} Taken together, these results place the diversity signal within origin-of-life and statistical-biology frameworks that emphasize network-level organization over molecular identity, and view biological systems as departures from equilibrium-like chemical distributions.\cite{england2013statistical} Its persistence across molecular classes supports its use as a biosignature rooted in organizational principles rather than specific compounds.

At the same time, the diversity framework complements prior organics-based life-detection approaches that distinguish biotic from abiotic materials using molecular identity, chirality, isotopes, or targeted pattern-based metrics.\citep{schwendner2022microbial, cleaves2023robust, buckner2024quantifying} In that context, amino acids and amphiphilic compounds have long been recognized as astrobiologically informative molecular classes.\citep{deamer1986role, deamer2002first} Here, the diversity framework shows that origin-diagnostic information can also be recovered from abundance structure alone. This distinction is useful in practice. Biosignature studies often contend with compositional sparsity, variable extraction protocols, and incomplete species overlap across datasets. By operating on relative abundance structure rather than compound identity, diversity metrics enable comparisons across chemically diverse and methodologically inconsistent samples, including cases where compound-by-compound matching is unreliable or requires imputation, as is often the case in planetary applications.\cite{chan2019deciphering}

\section*{Diversity as a framework for life detection} \label{sec: discussion}

The diversity approach offers a promising strategy for detecting life beyond Earth. Current biosignature detection efforts on Europa, Enceladus, and Mars are constrained by precisely these limitations. To date, no mission has included an instrument capable of assessing chirality in complex organics, and isotopic analysis is typically limited to small volatiles. On \textit{Europa Clipper},\cite{vance2023investigating} the \textit{MAss Spectrometer for Planetary EXploration} (MASPEX)\cite{waite2024maspex} will measure molecular abundances in gases in Europa’s exosphere or potential plumes, and the \textit{SUrface Dust Analyzer} (SUDA)\cite{kempf2025suda} will analyze surface-ejected particles, but neither can reliably resolve stereochemistry or compound-specific isotopes. Proposed instruments for Enceladus may overcome some of these limitations.\cite{mackenzie2022science, mousis2022moonraker} On Mars, the \textit{Sample Analysis at Mars} (SAM) instrument aboard the \textit{Curiosity} rover\cite{mahaffy2012sample} can detect volatile organics and measure isotopic ratios in simple species, but neither chirality nor complex molecular patterns. The delayed \textit{ExoMars} mission is expected to carry the first in situ chiral-capable instrument---\textit{Mars Organic Molecule Analyzer} (MOMA).\cite{goesmann2017mars}
In this context, diversity analysis offers a tractable, instrument-agnostic framework compatible with current and upcoming datasets that requires only relative abundances of molecules within a coherent molecular family. These can be measured by mass spectrometry, spectroscopy, electrophoresis, or other methods.\cite{chan2019deciphering}

However, planetary environments are often harsh, and organic molecules are often selectively degraded.\cite{pavlov2024radiolytic} To evaluate the durability of the diversity signal under such conditions, we simulated one degradation process, namely, radiolysis of biotic and abiotic amino-acid profiles in Europa’s near-surface ice, which was shown to be the main driver of degradation of organic compounds therein.\citep{yoffe2025fluorescent} We find that the degraded signal diverges from its original state but only briefly resembles an abiotic profile before becoming too sparse to classify, as shown in Fig.~\ref{europa_model}. The radiolytic degradation model is described in the Supplementary Information.
The statistical distinction between biotic and abiotic profiles persists across depths and timescales relevant to astrobiological exploration. This robustness highlights the method’s utility where preservation is uncertain and chemical complexity is reduced. Under such conditions, biosignatures may be difficult to detect, even in situ.

Beyond binary separation, the framework resolves internal variability within biotic samples. We observe a continuum of preservation states shaped by environment, diagenesis, and initial molecular composition (Fig.~\ref{fig: result_1D_mds_biotic}). Moderately degraded samples retain partial structure, whereas more altered profiles converge toward sparse, abiotic-like distributions (Fig.~\ref{fig: results_aa}b). In some cases, these differences reflect variation in original composition rather than degradation alone.\cite{king1972amino, saitta2024non} The signal, therefore, encodes a superposition of both biogenicity and the preservation trajectory. This superposition introduces an inherent limitation. The signal reflects distributional structure rather than molecular identity and carries no phylogenetic or compound-specific information. Interpretation requires contextual and complementary evidence. As with other biosignatures, it is not definitive on its own, but its generality and resilience to degradation make it a useful component of a multi-pronged search strategy.

In summary, amino-acid and fatty-acid assemblages carry an origin-diagnostic signal in their diversity structure. Across environments, biotic and abiotic samples occupy distinct regions of diversity space shaped by biosynthetic organization. Because the signal depends only on relative abundance structure, it can remain detectable even as molecular identities degrade. Diversity analysis thus provides an instrument-agnostic framework for assessing biogenicity in molecular datasets relevant to planetary exploration.

\begin{figure*}[t!]
\centering
\includegraphics[scale=1.]{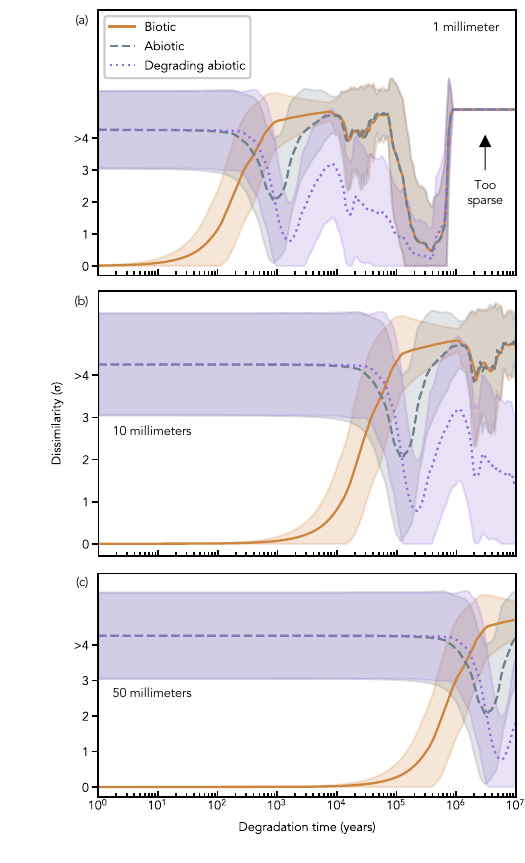}%
\caption{\textbf{Radiolytic degradation of the diversity signal of amino-acid profiles in Europa’s near-surface ice.}
Dissimilarity of the diversity signal for biotic and abiotic profiles on the leading hemisphere (60$^\circ$ latitude, 90$^\circ$W longitude).  
(\textbf{a}–\textbf{c}) Radiolytic degradation at depths of 1, 10, and 50~millimeters, respectively.  
Curves show the mean dissimilarity $z$-score of the degrading biotic evenness-curve distribution relative to three benchmarks:  
(\textbf{1}) the pristine biotic distribution (solid orange line),  
(\textbf{2}) the pristine abiotic distribution (dashed gray line), and  
(\textbf{3}) the simultaneously degrading abiotic distribution (dotted purple line).  
Shaded regions show one standard deviation margins across 100 simulations, each based on a randomly drawn biotic--abiotic sample pair and uncertainty-propagated evenness-curve ensembles. When radiolysis reduces absolute abundances to extremely low values such that an evenness curve is no longer defined, the corresponding intervals are labeled \textit{Too sparse}.}
\label{europa_model}
\end{figure*}

\bibliography{sample}

\section*{Methods}

\subsection*{Preprocessing}
\label{sec: preprocessing}

Amino-acid samples compiled in this study span a wide range of sampling strategies, extraction protocols, and quantification techniques. To ensure comparability across these diverse sources, we apply a consistent normalization to the ecodiversity metric. Because the evenness function, $E(q)$, is defined over relative abundances in each assemblage, all samples are treated as compositional vectors. This renders the need for scaling absolute measurement units unnecessary, and they remain unchanged. Specifically, for each sample, species-level abundances are normalized to obtain relative frequencies $p_j$, such that the vector $\mathbf{p}$ satisfies $\sum_j p_j = 1$. This normalization ensures that all samples are directly comparable within the $E(q)$ framework.

The term “sample” in this context refers to any compositionally coherent amino-acid profile, but its definition varies across the studies from which it was extracted. In some cases, such as with asteroidal or meteoritic extracts, a single sample represents a single amino-acid profile for that object.\cite{glavin1999nakhla, martins2007indigenous, Glavin2025} In other cases, multiple profiles exist from the same object under differing conditions (e.g., pristine and contaminated samples of the same meteorite), and they are retained as individual samples to reflect this distinction.\cite{glavin2021az} Where several profiles originate from a common context and exhibit compositional similarity, we aggregate them into a single averaged sample to improve representativeness.\cite{klevenz2010concentrations,fuchida2015concentrations, saitta2024non} This procedure is guided by metadata from each study. Table~1 summarizes the samples used in the analysis, including their provenance, treatment, and the logic of inclusion. For studies reporting separate concentrations of L- and D-enantiomers, we sum them to obtain total amino-acid abundances.

\subsection*{Richness and evenness of samples}
\label{sec: richness_and_evenness}

We use the term \textit{diversity} to describe the compositional structure of amino-acid mixtures across terrestrial and extraterrestrial settings. We treat the set of detected amino acids in each sample as an assemblage, analogous to an ecological community in which each compound corresponds to a species and its measured concentration represents its abundance. Diversity quantifies the structure of such assemblages. It reflects both the number of distinct components (richness) and their relative abundances (evenness). Richness increases with the number of detected compounds. Evenness is maximal when the compounds in the sample are uniformly distributed and minimal when the distribution is highly skewed, as illustrated in Fig.~\ref{fig: evenness_richness_illustration}.

For a given $i^{\rm{th}}$ sample, we quantify amino-acid assemblage diversity in its generalized form by computing Hill numbers, $D_q^{(i)}$, which incorporates both richness and evenness under a single parametric form.\cite{hill1973diversity} For a sample with $S_i$ detected compounds, and corresponding relative abundance of the $j^{\rm{th}}$ compound in the $i^{\rm{th}}$ sample, $p_{ij}$, the diversity of order $q \geq 0$ is defined as

\begin{equation}
D_q^{(i)} = \left( \sum_{j=1}^{S_i} p_{ij}^q \right)^{\frac{1}{1 - q}}.
\end{equation}

This expression interpolates between several diversity measures: for $q = 0$, $D_q$ equals the observed richness.\cite{magurran2013measuring} For $q = 1$, it converges to $\exp(H)$, where $H$ is the Shannon entropy and reflects the uncertainty in predicting a randomly drawn compound.\cite{shannon1948mathematical} For $q = 2$, it corresponds to the inverse Simpson index, which emphasizes the probability of repeated draws from dominant compounds.\cite{simpson1949measurement} Essentially, as $q$ increases, $D_q$ becomes increasingly sensitive to the most abundant species, suppressing the influence of rare ones.\cite{chao2014unifying} This framework enables diversity comparisons across samples with varying abundance distributions within a common statistical framework.

 For $q = 0$, $D_q$ equals richness. At $q = 1$, it reduces to $\exp(H)$, the exponential of Shannon entropy. Higher values of $q$ emphasize dominant species and suppress rare ones.

We derive evenness curves $E_i(q)$ by normalizing $D_q^{(i)}$ against its maximum for a given richness. Specifically,

\begin{equation} \label{eq_Eq}
E_i(q) = \frac{D_q^{(i)} - 1}{S_i - 1},
\end{equation}

where $S_i = D_0^{(i)}$ is the observed richness of the $i^{\rm{th}}$ sample. This formulation ensures that $E(q) \in [0, 1]$. Samples whose $E(q)$ values approach 1 contain uniformly distributed compounds, whereas samples with $E(q)$ values near 0 are dominated by a few compounds that are more abundant than others.

\subsection*{Uncertainty estimation and evenness curves distributions}
\label{sec: uncertainty_estimation}

Each reported abundance is accompanied by an uncertainty measure. This reflects variability introduced during extraction, quantification, and analysis, and is essential for constructing a statistically grounded representation of the sample. These uncertainties vary across studies and are often not explicitly reported, but they are inherent to all abundance measurements, regardless of the methodology. To account for this, we associate each sample with an explicit uncertainty model. This serves two purposes: First, it preserves the integrity of the measurement: abundance values are treated not as fixed quantities, but as estimates with finite precision. Second, it enables the controlled generation of synthetic compositions through drawing multiple realizations of abundance profiles from distributions informed by uncertainty. Without defined uncertainties, this process becomes arbitrary, and any structure inferred from the data loses statistical grounding. Uncertainty is not an auxiliary quantity, but a part of the data model.

Here, we consider three scenarios for assigning an uncertainty value to the $j^{\rm{th}}$ species in a sample: (1) When a single profile of a coherent molecular family (e.g., amino acids, fatty acids) is reported with per-species uncertainties, typically as standard deviations across replicate measurements. In this case, the noise is assumed to be additive, and we assign a normal distribution to each species abundance, centered on its reported value with the reported uncertainty as its standard deviation.\cite{carroll2006measurement} (2) When $K$ profiles are averaged into a single sample, and each has a reported uncertainty value, we propagate these as standard errors according to
\begin{equation}
\mathrm{SEM}_j = \frac{\sqrt{\sum_{k=1}^{K} \sigma_{j,k}^2}}{K},
\end{equation}
where $\sigma_{j,k}$ is the reported standard deviation of the measured abundance of the $j^{\rm{th}}$ species in the $k^{\rm{th}}$ profile. (3) When no uncertainties are reported, we estimate uncertainty by computing the empirical standard deviation of each species across the $K$ contributing profiles and derive the standard error as
\begin{equation}
\mathrm{SEM}_j = \frac{s_j}{\sqrt{K}},
\end{equation}
where $s_j$ is the empirical standard deviation of the $j^{\rm{th}}$ species across the averaged profiles, computed as
\[
s_j = \sqrt{\frac{1}{K - 1} \sum_{k=1}^{K} \left(x_{j,k} - \bar{x}_j\right)^2},
\]
with $x_{j,k}$ the abundance of the $j^{\rm{th}}$ species in the $k^{\rm{th}}$ profile, and $\bar{x}_j$ the corresponding sample mean. In this case, we assume a Student's $t$-distribution with $\nu = K - 1$ degrees of freedom. This choice rests on the assumption that the abundances of each species are approximately normally distributed across the averaged profiles. In this context, normality means that species-wise fluctuations in abundance across the averaged profiles are expected to be symmetric about the average and dominated by random variation rather than systematic bias. This is a standard approximation for estimating the uncertainty of a mean from a small sample.\cite{gelman1995bayesian, casella2024statistical} While the underlying distribution is not directly known, we mitigate this limitation by restricting such averaging to profiles that originate from a common experimental or environmental context, as reported in the source studies (see Preprocessing). This restriction limits variance to sources intrinsic to the sampling context, thereby avoiding, as much as possible, inflation from unrelated experimental or environmental differences.

This choice reflects both practical and empirical considerations: Most uncertainties are reported as symmetric errors around a mean, making the normal and $t$ distributions appropriate models. The $t$-distribution, in particular, accommodates broader uncertainty where the number of averaged profiles is low. While these distributions permit negative draws, such values are set to zero prior to normalization. This reflects the fact that abundances near detection limits may plausibly vanish within error, thereby avoiding the imposition of artificial bounds.

Lastly, when measurement errors are explicitly reported as relative (i.e., proportional to abundance), we consider the log-normal distribution a more appropriate model. This reflects the fact that multiplicative variability induces asymmetry in the error structure that is better captured in log space. When the relative uncertainty is estimated from averaged profiles, we adopt the log-$t$ distribution, which preserves the multiplicative structure while accounting for uncertainty in variance estimation.

With this schema, each sample is represented by a set of parametric distributions over species abundances: normally distributed when measurement uncertainty is reported directly, $t$-distributed when it is inferred from the variation among averaged profiles, log-normally distributed when the reported uncertainty is explicitly relative, and log-$t$-distributed when uncertainty is relative and inferred from the data. This enables consistent propagation of uncertainty across the diversity framework while respecting differences in data quality and availability. The end result is that in each sample, each species is assigned an uncertainty value $\epsilon_j$, derived from one of the expressions above, and treated as the scale parameter of its associated distribution.

To generate the distribution of evenness curves, we treat each sample as a distribution over plausible compositions. For each $j^{\rm{th}}$ compound in the $i^{\rm{th}}$ sample, the reported abundance $\mu_{ij}$ and its associated uncertainty $\epsilon_{ij}$ define a parametric distribution from which absolute concentrations are drawn. In most cases, we sample from a normal distribution, $x_{ij} \sim \mathcal{N}(\mu_{ij}, \epsilon_{ij}^2)$; for low-confidence estimates or heavy-tailed uncertainties, we instead draw from a Student’s $t$-distribution: $x_{ij} \sim \mu_{ij} + \epsilon_{ij} \cdot t_{\nu_{ij}}$. In cases where measurement errors are explicitly relative, we sample from a log-normal distribution, 
\[
\log x_{ij} \sim \mathcal{N} \left( \log \mu_{ij}, \left( \frac{\epsilon_{ij}}{\mu_{ij}} \right)^2 \right),
\]
or, when the relative uncertainty is inferred from the data, from a log-$t$ distribution,
\[
\log x_{ij} \sim t_{\nu_{ij}} \left( \log \mu_{ij}, \frac{\epsilon_{ij}}{\mu_{ij}} \right).
\]
and exponentiate the result. Each abundance vector is then normalized to yield relative frequencies $p_{ij} = x_{i,\ell=j} / \sum_\ell x_{i\ell}$, and transformed into an evenness curve using equation~(\ref{eq_Eq}), across a $q$-domain. The resulting empirical distribution $\left\{ E_i^{(n)}(q) \right\}_{n=1}^N$ captures the propagation of measurement uncertainty through the diversity formalism under the appropriate sampling regime. A detailed account of the statistical model assigned to each sample, the specific profiles used, and the rationale for their selection is provided in the Supplementary Information.

\subsection*{Dissimilarity between samples}
\label{sec: dissimilarity}

Several metrics have been proposed for quantifying the similarity between evenness curves. Prior studies have employed a range of approaches, including pointwise permutation tests across selected values of $q$, comparisons based on area under the curve (AUC), and various metrics derived from Functional Data Analysis (FDA).\cite{di2016environmental, vishne2025diversity, golini2025functional} However, evenness curves are, by construction, smooth and monotonically decreasing functions of $q$, with values inherently coupled through their shared dependence on the underlying abundance distribution. Consequently, FDA-based techniques exaggerate small global displacements by accumulating their effects across the domain of $q$. This results in separations with inflated statistical significance even when compositional differences are small. A detailed comparison of several dissimilarity metrics is provided in the Supplementary Information. 

Moreover, distributions of evenness curves are nonlinearly dependent on the underlying distributions of species abundances from which they are computed (see Richness and Evenness of Samples). This precludes the use of parametric significance tests, such as $z$- or $t$-tests, and requires a nonparametric approach that treats the distributions empirically.\cite{efron1994introduction}

To address this, we adopt a nonparametric framework to compute a rank-based, two-sided empirical $p$-value:\cite{lehmann1986testing} At each pair of samples, we consider their evenness curves distributions $\{ E^{(n)}_1(q) \}_{n=1}^N$ and $\{ E^{(n)}_2(q) \}_{n=1}^N$. For each discretized diversity order $q_j > 0$, we evaluate the degree of directional overlap between the two distributions as

\begin{equation} \label{eq: p_val}
p_{\mathrm{emp}}(q_j) = \frac{2}{N^2} \min \left( \sum_{m,n} \mathbb{I}\left[ E^{(m)}_1(q_j) > E^{(n)}_2(q_j) \right], \sum_{m,n} \mathbb{I}\left[ E^{(m)}_1(q_j) < E^{(n)}_2(q_j) \right] \right),
\end{equation}

where $\mathbb{I}[\cdot]$ is the indicator function. This statistic quantifies the directional imbalance between the two ensembles without requiring parametric assumptions about the shape or variance of the underlying distributions. Evaluation at $q = 0$ is skipped because evenness is undefined at this value: $E(0) = (D_0 - 1)/(S - 1) = 1$ for any distribution with support size $S$, causing any pair of curves to overlap and reducing discriminatory power. The minimum operation ensures a conservative two-sided estimate, bounded below by $2/N^2$ to avoid zero-valued results in finite samples.

To ensure robustness against sampling noise, we apply a Wilson score correction to each $p_{\mathrm{emp}}(q_j)$, retaining the upper bound of its confidence interval at $\alpha = 0.05$. This yields a conservative estimate of the minimal separability between distributions, defined as the maximal corrected $p$-value across all $q$:

\begin{equation} \label{eq: wilson}
p_{\max} = \max_j \, \mathrm{WilsonUpper}\left(p_{\mathrm{emp}}(q_j) \mid \alpha = 0.05\right).
\end{equation}

To control for multiple comparisons between all sample pairs, we apply a Benjamini--Hochberg false discovery rate (FDR) correction \cite{benjamini1995controlling} to the set of Wilson-corrected $p$-values.
This procedure identifies the point of weakest statistical separation between the two evenness distributions. If the samples are meaningfully distinct, they must differ across a substantial portion of the $q$ domain. By reporting the maximal corrected $p$-value, this method avoids inflating separability from cumulative trends and emphasizes robust distinctions in evenness structure. To express this dissimilarity on a standardized scale, we convert the FDR-corrected $p_{\max}$ to a $z$-score by inverting the survival function of the standard normal distribution:

\begin{equation} \label{eq: z-score}
z = \Phi^{-1}(1 - p_{\max}^{(\rm{FDR})}),
\end{equation}

where $\Phi^{-1}$ denotes the inverse cumulative distribution function (quantile function) of the standard normal. 

\subsection*{Data availability}

All data used in this study are available at Zenodo.\citep{yoffe2026_molecular_diversity_zenodo}

\subsection*{Code availability}

The code used in this study is available at Zenodo.\citep{yoffe2026_molecular_diversity_zenodo}

\section*{Acknowledgements}

We acknowledge Robert Colwell, Anne Chao, Shay Zucker, and Chris McKay for contributing to this work through useful discussions. 
We thank the two anonymous reviewers for their comments, which substantially improved the scope and quality of this work.
F.K. acknowledges support from startup funds from the University of California, Riverside.

\begin{appendices}

\begin{figure*}[b!]
\centering
\rotatebox[origin=c]{0}{\includegraphics[scale = 0.8]{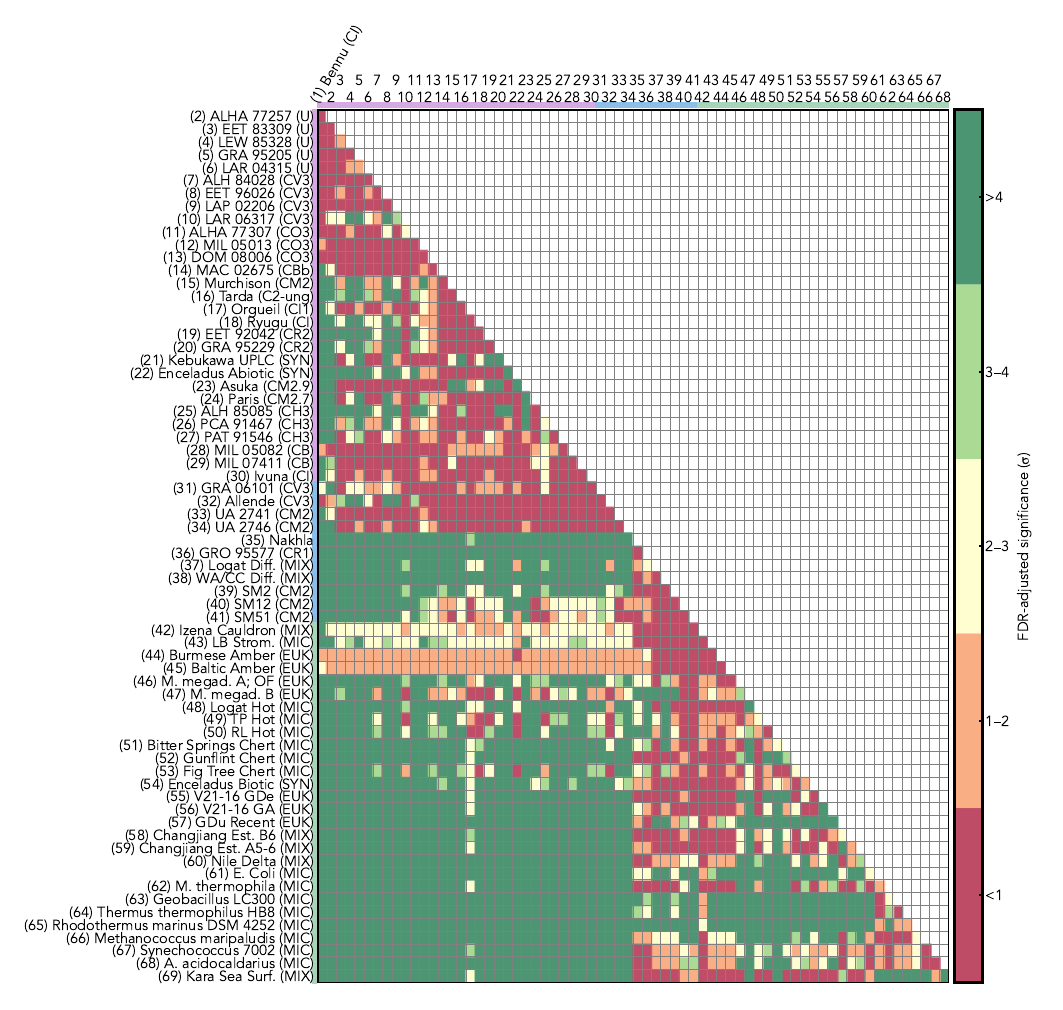}}
  \caption{\textbf{Pairwise dissimilarity matrix between amino-acid assemblages.} Each cell quantifies the separation between the evenness curve distributions of two samples, expressed and color-coded in standard deviations ($\sigma$), derived from the corrected $p$-values. The colors adjacent to the sample names denote their inferred origin: pink indicates abiotic, blue indicates mixed, and green indicates biotic.}
  \label{fig: z_matrix_aa}
\end{figure*}

\begin{figure*}
\centering
\rotatebox[origin=c]{0}{\includegraphics[scale = 0.8]{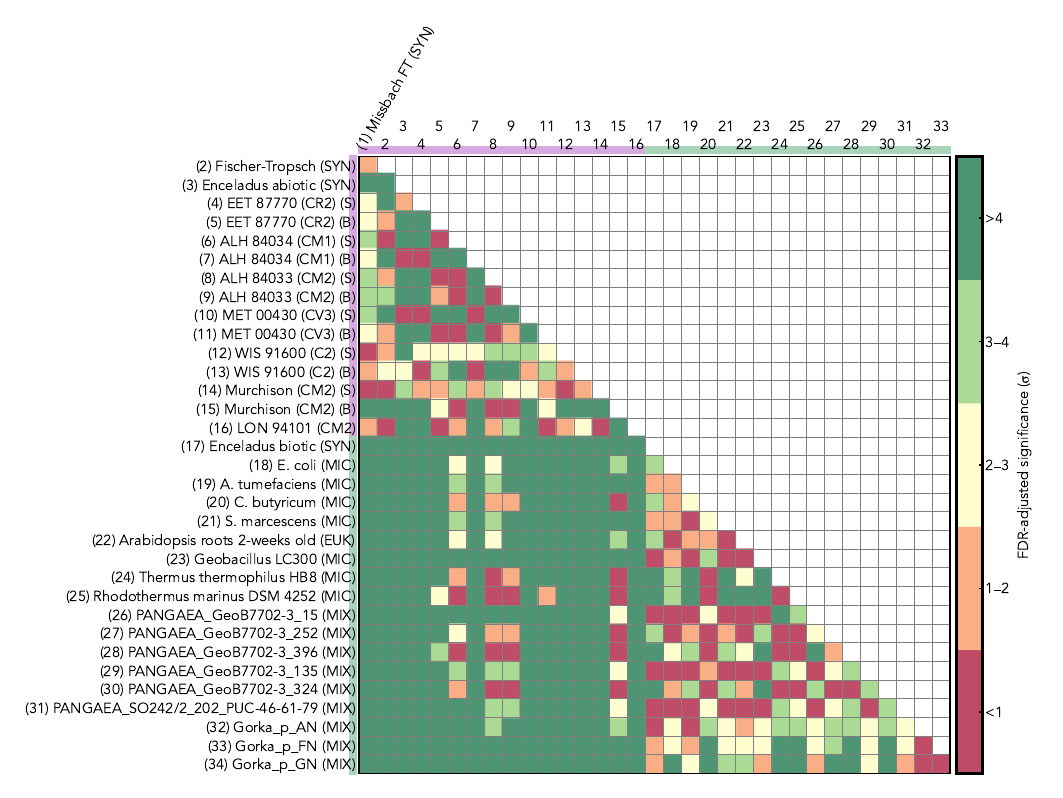}}
  \caption{\textbf{Pairwise dissimilarity matrix between fatty-acid assemblages.} Each cell quantifies the separation between the evenness curve distributions of two samples, expressed and color-coded in standard deviations ($\sigma$), similar in format to Extended~Data Figure~\ref{fig: z_matrix_aa}.}
  \label{fig: z_matrix_fa}
\end{figure*}

\end{appendices}

\FloatBarrier

\begin{supplementary}

\section{Sample description tables}

\begin{sidewaystable}[h!]
\tiny
\centering
\caption{\textbf{Summary of amino-acid assemblages included in this work.} Each entry specifies the sample name, the source object, the inferred biogenic or abiotic origin, and the data provenance.}
\label{tab:database_table}
\begin{tabular}{lllll}
\hline
\textbf{Index} & \textbf{Sample Name} & \textbf{Sample Object} & \textbf{Inferred Origin} & \textbf{Provenance} \\
\hline

1 & Bennu (CI) & Asteroid Bennu (B-type), space & Abiotic & \cite{Glavin2025} \\
2 & Murchison (CM2) & Murchison meteorite (CM2) (Australia; fell in 1969) & Abiotic & \cite{Glavin2025} \\
3 & Tarda (C2) & Tarda meteorite (C2) (Morocco; found in 2020) & Abiotic & \cite{Glavin2025} \\
4 & Orgueil (CI1) & Orgueil meteorite (CI1) (France; fell in 1864) & Abiotic & \cite{glavin2010effects} \\
5 & Ryugu (CI) & Asteroid Ryugu (C-type), space & Abiotic & \cite{parker2023extraterrestrial} \\
6 & Asuka (CM2.9) & Asuka~12236 meteorite (CM2) (Antarctica; found in 2012) & Abiotic & \cite{glavin2020asuka} \\
7 & EET~92042 (CR2) & Elephant Moraine 92042 meteorite (CR2) (Antarctica; found in 1992) & Abiotic & \cite{martins2007indigenous} \\
8 & GRA~95229 (CR2) & Graves Nunataks 95229 meteorite (CR2) (Antarctica; found in 1995) & Abiotic & \cite{martins2007indigenous} \\
9 & Kebukawa~UPLC (SYN) & Amino acids synthesized in a laboratory hydrothermal reactor (150$^\circ$C) & Abiotic & \cite{kebukawa2017one} \\
10 & Enceladus Abiotic (SYN) & Simulated abiotic abundances for the Enceladus ocean & Abiotic & \cite{klenner2020discriminating} \\
11 & Logat~Diff (MIX) & Sedimentary organics near diffuse hydrothermal systems (Logatchev field; $\sim$20$^\circ$C) & Mixed & \cite{klevenz2010concentrations} \\
12 & CC/WA~Diff (MIX) & Sedimentary organics near diffuse hydrothermal systems (Wideawake/Comfortless Cove fields; $\sim$20$^\circ$C) & Mixed & \cite{klevenz2010concentrations} \\
13 & UA~2741 (CM2) & Pre-rainfall fragment of the Aguas~Zarcas meteorite (CM2) & Mixed (abiotic + biotic contamination) & \cite{glavin2021az} \\
14 & UA~2746 (CM2) & Pre-rainfall fragment of the Aguas~Zarcas meteorite (CM2) & Mixed (abiotic + biotic contamination) & \cite{glavin2021az} \\
15 & GRO~95577 (CR1) & Grosvenor Mountains 95577 meteorite (CR1) (Antarctica; found in 1995) & Mixed (abiotic + biotic contamination) & \cite{martins2007indigenous} \\
16 & Nakhla & Nakhla meteorite (Martian nakhlite) (Egypt; found in 1911) & Mixed (abiotic + biotic contamination) & \cite{glavin1999nakhla} \\
17 & Nile~Delta (MIX) & Nearshore marine sediment core (0--15 cm) from the Nile Delta ($\sim$120 m depth; collected 1964) & Biotic & \cite{glavin1999nakhla} \\
18 & RL~Hot (MIC) & Red Lion black smoker hydrothermal fluids ($\sim$350$^\circ$C) & Biotic & \cite{klevenz2010concentrations} \\
19 & TP~Hot (MIC) & Turtle Pits black smoker hydrothermal fluids ($\sim$340--350$^\circ$C) & Biotic & \cite{klevenz2010concentrations} \\
20 & Logat~Hot (MIC) & Logatchev black smoker hydrothermal fluids ($\sim$340--350$^\circ$C) & Biotic & \cite{klevenz2010concentrations} \\
21 & Izena~Cauldron (MIX) & Hydrothermally influenced seafloor sediment from the Izena Cauldron (Okinawa Trough) & Biotic & \cite{fuchida2015concentrations} \\
22 & Bitter~Springs~Chert (MIC) & Precambrian black chert ($\sim$1.0 Ga) from the Bitter Springs Formation (Australia) & Biotic (marine microbial) & \cite{schopf1968amino} \\
23 & Gunflint~Chert (MIC) & Precambrian chert ($\sim$1.9 Ga) from the Gunflint Iron Formation (Ontario, Canada) & Biotic (microfossil-bearing) & \cite{schopf1968amino} \\
24 & Fig~Tree~Chert (MIC) & Archean black chert ($>$3.1 Ga) from the Fig Tree Group (South Africa) & Possibly biotic, highly degraded & \cite{schopf1968amino} \\
25 & Enceladus Biotic (SYN) & Simulated biotic abundances for the Enceladus ocean & Biotic & \cite{klenner2020discriminating} \\
26 & M.~megad.~A;~OF (EUK) & \textit{Megaloolithus megadermus} titanosaur eggshell (outer flakes THAA; Late Cretaceous, Argentina) & Biotic & \cite{saitta2024non} \\
27 & M.~megad.~B (EUK) & \textit{Megaloolithus megadermus} titanosaur eggshell (whole shell, intra-crystalline THAA; Late Cretaceous, Argentina) & Biotic & \cite{saitta2024non} \\
28 & Baltic~Amber (EUK) & Fossil feather preserved in Baltic amber ($\sim$44 Ma; Eocene, Northern Europe) & Biotic & \cite{mccoy2019ancient} \\
29 & Burmese~Amber (EUK) & Fossil feather preserved in Burmese amber ($\sim$99 Ma; Cretaceous, Myanmar) & Biotic & \cite{mccoy2019ancient} \\
30 & V21-16~GDe (EUK) & Fossil foraminifer (\textit{Globoquadrina dehiscens}) (Early Miocene, $\sim$18 Ma) & Biotic (marine plankton) & \cite{king1972amino} \\
31 & V21-16~GA (EUK) & Fossil foraminifer (\textit{Globoquadrina altispira}) (Early Miocene, $\sim$18 Ma) & Biotic (marine plankton) & \cite{king1972amino} \\
32 & GDu~Recent (EUK) & Modern foraminifer (\textit{Globoquadrina dutertrei}), Recent marine & Biotic (marine plankton) & \cite{king1972amino} \\
33 & Changjiang~Est.~B6 (MIX) & Surface sediment core from nearshore station (Changjiang Estuary; $\sim$8 m water depth) & Biotic (terrestrial + marine, less degraded) & \cite{wei2022variability} \\
34 & Changjiang~Est.~A5-6 (MIX) & Surface sediment core from offshore station (Changjiang Estuary; $\sim$46 m water depth) & Biotic (terrestrial + marine, more degraded) & \cite{wei2022variability} \\
35 & LB~Strom. (MIC) & Jurassic stromatolitic layers from the \textit{Rosso Ammonitico Veronese} (Italy; $\sim$160--170 Ma) & Biotic (cyanobacterial + fungal) & \cite{ballarini1994amino} \\

\hline
\end{tabular}
\end{sidewaystable}

\begin{sidewaystable}[h!]
\tiny
\centering
\ContinuedFloat
\caption[]{Summary of amino acid assemblages included in this work (continued). Each entry specifies the sample name, the source object, the inferred biogenic or abiotic origin, and the data provenance.}
\begin{tabular}{lllll}
\hline
\textbf{Index} & \textbf{Sample Name} & \textbf{Sample Object} & \textbf{Inferred Origin} & \textbf{Provenance} \\
\hline

36 & Paris (CM2.7) & CM2.7 carbonaceous chondrite amino acids & Abiotic & \cite{martins2015amino} \\
37 & ALH~85085 (CH3) & CH3 carbonaceous chondrite (Antarctica), analyzed for indigenous amino acid abundances & Abiotic & \cite{burton2013extraterrestrial} \\
38 & PCA~91467 (CH3) & CH3 carbonaceous chondrite (Antarctica), analyzed for indigenous amino acid abundances & Abiotic & \cite{burton2013extraterrestrial} \\
39 & PAT~91546 (CH3) & CH3 carbonaceous chondrite (Antarctica), analyzed for indigenous amino acid abundances & Abiotic & \cite{burton2013extraterrestrial} \\
40 & MAC~02675 (CBb) & CBb carbonaceous chondrite (Antarctica), analyzed for indigenous amino acid abundances & Abiotic & \cite{burton2013extraterrestrial} \\
41 & MIL~05082 (CB) & CB carbonaceous chondrite (Antarctica), analyzed for indigenous amino acid abundances & Abiotic & \cite{burton2013extraterrestrial} \\
42 & MIL~07411 (CB) & CB carbonaceous chondrite (Antarctica), analyzed for indigenous amino acid abundances & Abiotic & \cite{burton2013extraterrestrial} \\
43 & ALHA~77257 (U) & Antarctic ureilite amino acids & Abiotic & \cite{burton2012propensity} \\
44 & EET~83309 (U) & Antarctic ureilite amino acids & Abiotic & \cite{burton2012propensity} \\
45 & LEW~85328 (U) & Antarctic ureilite amino acids & Abiotic & \cite{burton2012propensity} \\
46 & GRA~95205 (U) & Antarctic ureilite amino acids & Abiotic & \cite{burton2012propensity} \\
47 & LAR~04315 (U) & Antarctic ureilite amino acids & Abiotic & \cite{burton2012propensity} \\
48 & ALH~84028 (CV3) & CV3 carbonaceous chondrite amino acids & Abiotic & \cite{burton2012propensity} \\
49 & EET~96026 (CV3) & CV3 carbonaceous chondrite amino acids & Abiotic & \cite{burton2012propensity} \\
50 & LAP~02206 (CV3) & CV3 carbonaceous chondrite amino acids & Abiotic & \cite{burton2012propensity} \\
51 & GRA~06101 (CV3) & CV3 carbonaceous chondrite amino acids & Abiotic & \cite{burton2012propensity} \\
52 & LAR~06317 (CV3) & CV3 carbonaceous chondrite amino acids & Abiotic & \cite{burton2012propensity} \\
53 & ALHA~77307 (CO3) & CO3 carbonaceous chondrite amino acids & Abiotic & \cite{burton2012propensity} \\
54 & MIL~05013 (CO3) & CO3 carbonaceous chondrite amino acids & Abiotic & \cite{burton2012propensity} \\
55 & DOM~08006 (CO3) & CO3 carbonaceous chondrite amino acids & Abiotic & \cite{burton2012propensity} \\
56 & Allende (CV3) & CV3 carbonaceous chondrite amino acids & Mixed (extraterrestrial contaminated) & \cite{burton2012propensity} \\
57 & Ivuna (CI) & CI carbonaceous chondrite amino acids & Abiotic & \cite{burton2014effects} \\
58 & SM2 (CM2) & CM2 carbonaceous chondrite amino acids & Mixed (extraterrestrial contaminated) & \cite{burton2014effects} \\
59 & SM12 (CM2) & CM2 carbonaceous chondrite amino acids & Mixed (extraterrestrial contaminated) & \cite{burton2014effects} \\
60 & SM51 (CM2) & CM2 carbonaceous chondrite amino acids & Mixed (extraterrestrial contaminated) & \cite{burton2014effects} \\
61 & \textit{E.~coli} K-12 MG1655 (MIC) & Laboratory-grown bacterial biomass & Biotic (bacterial) & \cite{simensen2022experimental} \\
62 & \textit{M.~thermophila} (EUK) & Filamentous fungal biomass & Biotic (fungal) & \cite{jiang2024} \\
63 & Geobacillus sp.~LC300 (MIC) & Extremely thermophilic bacterial biomass & Biotic (bacterial) & \cite{cordova201713c} \\
64 & \textit{Thermus thermophilus}~HB8 (MIC) & Extremely thermophilic bacterial biomass & Biotic (bacterial) & \cite{cordova201713c} \\
65 & \textit{Rhodothermus marinus}~DSM~4252 (MIC) & Extremely thermophilic bacterial biomass & Biotic (bacterial) & \cite{cordova201713c} \\
66 & \textit{Methanococcus maripaludis} (MIC) & Methanogenic archaeal single-cell protein biomass grown in microbial electrolysis cells & Biotic (archaeal) & \cite{wang2024microbial} \\
67 & \textit{Synechococcus} sp.~PCC~7002 (MIC) & Laboratory-grown cyanobacterial biomass composition & Biotic (cyanobacterial) & \cite{beck2018measuring} \\
68 & \textit{Alicyclobacillus acidocaldarius}~DSM~446 (MIC) & Thermophilic bacterial biomass composition & Biotic (bacterial) & \cite{beck2018measuring} \\
69 & Kara Sea Surf. (MIX) & Suspended particulate matter amino-acid composition (surface waters, Kara Sea) & Biotic (marine SPM) & \cite{gaye2022aaah} \\

\hline
\end{tabular}
\end{sidewaystable}

\begin{sidewaystable}[h!]
\tiny
\centering
\caption{\textbf{Summary of fatty-acid assemblages included in this work.} Each entry specifies the sample name, source object, inferred biogenic or abiotic origin, and the literature reference for the data used.}
\label{tab: database_table_fatty}
\begin{tabular}{lllll}
\hline
 & \textbf{Sample Name} & \textbf{Sample Object} & \textbf{Inferred Origin} & \textbf{Provenance} \\
\hline

1 & Fischer-Tropsch (SYN) & Typical fatty-acid profile resulting from laboratory Fischer-Tropsch synthesis & Abiotic & \cite{mccollom2007abiotic} \\

2 & Enceladus abiotic (SYN) & Simulated abiotic abundances for Enceladus's ocean & Abiotic & \cite{klenner2020analog} \\

3 & Enceladus biotic (SYN) & Simulated biotic abundances for Enceladus's ocean & Biotic & \cite{klenner2020discriminating} \\

4 & Missbach~FT (SYN) & Fatty-acid distribution produced by laboratory Fischer--Tropsch--type synthesis & Abiotic & \cite{missbach2018assessing} \\

5 & EET 87770 (CR2) & Carbonaceous chondrite (CR2), meteoritic fatty-acid abundances & Abiotic & \cite{aponte2011effects} \\

6 & ALH 84034 (CM1) & Carbonaceous chondrite (CM1), aqueously altered meteoritic fatty acids & Abiotic & \cite{aponte2011effects} \\

7 & ALH 84033 (CM2) & Carbonaceous chondrite (CM2), meteoritic fatty-acid abundances & Abiotic & \cite{aponte2011effects} \\

8 & MET 00430 (CV3) & Carbonaceous chondrite (CV3), minimally altered meteoritic fatty acids & Abiotic & \cite{aponte2011effects} \\

9 & WIS 91600 (C2) & Carbonaceous chondrite (C2), meteoritic fatty-acid abundances & Abiotic & \cite{aponte2011effects} \\

10 & Murchison (CM2) & Carbonaceous chondrite (CM2), meteoritic fatty-acid abundances & Abiotic & \cite{aponte2011effects} \\

11 & LON 94101 (CM2) & Carbonaceous chondrite (CM2), meteoritic fatty-acid abundances & Abiotic & \cite{aponte2014chirality} \\

12 & Geobacillus LC300 (MIC) & Phospholipid-derived fatty acids from thermophilic bacterium & Biotic & \cite{cordova201713c} \\

13 & Thermus thermophilus HB8 (MIC) & Phospholipid-derived fatty acids from thermophilic bacterium & Biotic & \cite{cordova201713c} \\

14 & Rhodothermus marinus DSM 4252 (MIC) & Phospholipid-derived fatty acids from thermophilic bacterium & Biotic & \cite{cordova201713c} \\

15 & E. coli (MIC) & Phospholipids of \emph{Escherichia coli} detected using gas-liquid chromatography & Biotic & \cite{hildebrand1964fatty} \\

16 & A. tumefaciens (MIC) & Phospholipids of \emph{Agrobacterium tumefaciens} detected using gas-liquid chromatography & Biotic & \cite{hildebrand1964fatty} \\

17 & C. butyricum (MIC) & Phospholipids of \emph{Clostridium butyricum} detected using gas-liquid chromatography & Biotic & \cite{hildebrand1964fatty} \\

18 & S. marcescens (MIC) & Phospholipids of \emph{Serratia marcescens} detected using gas-liquid chromatography & Biotic & \cite{hildebrand1964fatty} \\

19 & Arabidopsis roots 2-weeks old (PLA) & Two-week-old roots of \emph{Arabidopsis thaliana} & Biotic & \cite{delude2016primary} \\

20 & PANGAEA\_GeoB7702-3\_15 (MIX) &
Marine sediment core GeoB7702-3, eastern Mediterranean; fatty-acid assemblage at 15\,cm burial depth &
Mixed biogenic &
\cite{meyer2024rdbc} \\

21 & PANGAEA\_GeoB7702-3\_135 (MIX) &
Marine sediment core GeoB7702-3, eastern Mediterranean; fatty-acid assemblage at 135\,cm burial depth &
Mixed biogenic &
\cite{meyer2024rdbc} \\

22 & PANGAEA\_GeoB7702-3\_252 (MIX) &
Marine sediment core GeoB7702-3, eastern Mediterranean; fatty-acid assemblage at 252\,cm burial depth &
Mixed biogenic &
\cite{meyer2024rdbc} \\

23 & PANGAEA\_GeoB7702-3\_324 (MIX) &
Marine sediment core GeoB7702-3, eastern Mediterranean; fatty-acid assemblage at 324\,cm burial depth &
Mixed biogenic &
\cite{meyer2024rdbc} \\

24 & PANGAEA\_GeoB7702-3\_396 (MIX) &
Marine sediment core GeoB7702-3, eastern Mediterranean; fatty-acid assemblage at 396\,cm burial depth &
Mixed biogenic &
\cite{meyer2024rdbc} \\

25 & PANGAEA\_SO242/2\_202\_PUC-46-61-79 (MIC) &
Phospholipid-derived fatty acids (PLFAs) from RV \emph{SONNE} cruise SO242/2 push core &
Biotic &
\cite{hoving2022phospholipid} \\

 &  &
202\_PUC-46-61-79; sediment slice 2--5\,cm below seafloor (DISCOL site, Peru Basin, SE Pacific) &
 &
 \\

26 & Gorka\_p\_AN (MIX) &
Phospholipid-derived fatty-acid assemblage from agricultural soil collected in Lower Austria (Austria) &
Mixed biogenic &
\cite{gorka2023beyond} \\

27 & Gorka\_p\_FN (MIX) &
Phospholipid-derived fatty-acid assemblage from temperate beech forest soil collected in Lower Austria (Austria) &
Mixed biogenic &
\cite{gorka2023beyond} \\

28 & Gorka\_p\_GN (MIX) &
Phospholipid-derived fatty-acid assemblage from temperate grassland soil collected in Austria &
Mixed biogenic &
\cite{gorka2023beyond} \\

\hline
\end{tabular}
\end{sidewaystable}

\begin{sidewaystable}[h]
\tiny
\centering
\caption{\textbf{Samples used and uncertainty model applied for each sample in the amino-acid dataset.}}
\label{tab:samples_uncertainty}
\begin{tabular}{clll}
\hline
\textbf{Index} & \textbf{Sample Name} & \textbf{Samples Used} & \textbf{Uncertainty Model} \\
\hline

1 & Bennu (CI) \cite{Glavin2025} & - & Normal; Root Sum of Squares (RSS) of L and D (their Extended Table~3) \\
2 & Murchison (CM2) \cite{Glavin2025} & - & Normal; RSS of L and D (their Extended Table~3) \\
3 & Tarda (C2-ung) \cite{Glavin2025} & - & Normal; RSS of L and D (their Extended Table~3) \\
4 & Orgueil (CI1) \cite{glavin2010effects} & - & Normal; RSS of L and D (their Extended Table~3) \\
5 & Ryugu (CI) \cite{parker2023extraterrestrial} & - & Normal; RSS of L and D (their Extended Table~3) \\
6 & Asuka (CM2.9) \cite{glavin2020asuka} & - & Normal; RSS of L and D (their Tab.~2) \\
7 & EET~92042 (CR2) \cite{martins2007indigenous} & - & Normal; RSS of L and D (their Tab.~1) \\
8 & GRA~95229 (CR2) \cite{martins2007indigenous} & - & Normal; RSS of L and D (Tab.~1) \\
9 & Kebukawa~UPLC (SYN) \cite{kebukawa2017one} & - & Normal; RSS of L and D (their Tab.~1) \\
10 & Enceladus Abiotic (SYN) \cite{klenner2020discriminating} & - & None \\
11 & Logat Diff (MIX) \cite{klevenz2010concentrations} & Averaged over samples 8–9 & log-normal; Relative 10\% uncertainty of measured value of each profile \\
12 & CC/WA Diff (MIX) \cite{klevenz2010concentrations} & Averaged over samples 18,19, and 22 & log-normal; Relative 10\% uncertainty of measured value of each profile \\
13 & UA~2741 (CM2) \cite{glavin2021az} & - & Normal; RSS of L and D (their Tab.~1) \\
14 & UA~2746 (CM2) \cite{glavin2021az} & - & Normal; RSS of L and D (their Tab.~1) \\
15 & GRO~95577 (CR1) \cite{martins2007indigenous} & - & Normal; RSS of L and D (their Tab.~1) \\
16 & Nakhla \cite{glavin1999nakhla} & - & Normal; RSS of L and D (their Tab.~1) \\
17 & Nile~Delta (MIX) \cite{glavin1999nakhla}  & - & Normal; RSS of L and D (their Tab.~1) \\
18 & RL~Hot (MIC) \cite{klevenz2010concentrations} & Sample 27 & log-normal; Relative 10\% uncertainty of measured value \\
19 & TP~Hot (MIC) \cite{klevenz2010concentrations} & Averaged over samples 23–25 & log-normal; Relative 10\% uncertainty of measured value of each profile \\
20 & Logat~Hot (MIC) \cite{klevenz2010concentrations} & Averaged over samples 4,6,10,11,14, and 15 & log-normal; Relative 10\% uncertainty of measured value of each profile \\
21 & Izena~Cauldron (MIX) \cite{fuchida2015concentrations} & Averaged over samples PC-1,3,4 & $t$-dist.; Standard deviation inferred from the data \\
22 & Bitter~Springs~Chert (MIC) \cite{schopf1968amino} & - & log-normal; Reported relative uncertainty of 7\% of the reported values \\
23 & Gunflint~Chert (MIC) \cite{schopf1968amino} & - & log-normal; Reported relative uncertainty of 8\% of the reported values \\
24 & Fig~Tree~Chert (MIC) \cite{schopf1968amino} & - & log-normal; Reported relative uncertainty of 10\% of the reported values \\
25 & Enceladus Biotic (SYN) & - & None \\
26 & M. megad A; OF (EUK) \cite{saitta2024non} & Averaged over 4 samples tagged 12092bH* (their Tab.~S1) & $t$-dist.; Standard deviation inferred from the data \\
27 & M. megad B (EUK) \cite{saitta2024non} & Averaged over 3 samples tagged 12093bH* (their Tab.~S3) & $t$-dist.; Standard deviation inferred from the data \\
28 & Baltic Amber (EUK) \cite{mccoy2019ancient} & - & log-$t$; Standard deviation inferred from 3 samples of a chicken feather at 140°C after 3 weeks \\
29 & Burmese Amber (EUK) \cite{mccoy2019ancient} & - & log-$t$; Standard deviation inferred from 3 samples of a chicken feather at 140°C after 3 weeks \\
30 & V21-16 GDe (EUK) \cite{king1972amino} & - & log-normal; Reported relative uncertainty of 3\% of the reported values \\
31 & V21-16 GA (EUK) \cite{king1972amino} & - & log-normal; Reported relative uncertainty of 3\% of the reported values \\
32 & GDu Recent (EUK) \cite{king1972amino} & - & log-normal; Reported relative uncertainty of 3\% of the reported values \\
33 & Changjiang Est. B6 (MIX) \cite{wei2022variability} & $t$-dist; Averaged over all depths of sample B6 \\
34 & Changjiang Est. A5-6 (MIX) \cite{wei2022variability} & $t$-dist; Averaged over all depths of sample A5-6 \\
35 & LB Strom. (MIC) \cite{ballarini1994amino} & Averaged over samples LB-17,21, and 25 & $t$-dist.; Standard deviation inferred from the data \\

\hline
\end{tabular}
\end{sidewaystable}

\begin{sidewaystable}[h]
\tiny
\centering
\ContinuedFloat
\caption[]{\textbf{Samples used and uncertainty model applied for each sample in the amino-acid dataset (continued).}}
\label{tab:samples_uncertainty2}
\begin{tabular}{clll}
\hline
\textbf{Index} & \textbf{Sample Name} & \textbf{Samples Used} & \textbf{Uncertainty Model} \\
\hline

36 & Paris (CM2.7) \cite{martins2015amino} & - & Normal; RSS of L and D \\
37 & ALH~85085 (CH3) \cite{burton2013extraterrestrial} & - & Normal; RSS of L and D \\
38 & PCA~91467 (CH3) \cite{burton2013extraterrestrial} & - & Normal; RSS of L and D \\
39 & PAT~91546 (CH3) \cite{burton2013extraterrestrial} & - & Normal; RSS of L and D \\
40 & MAC~02675 (CBb) \cite{burton2013extraterrestrial} & - & Normal; RSS of L and D \\
41 & MIL~05082 (CB) \cite{burton2013extraterrestrial} & - & Normal; RSS of L and D \\
42 & MIL~07411 (CB) \cite{burton2013extraterrestrial} & - & Normal; RSS of L and D \\
43 & ALHA~77257 (U) \cite{burton2012propensity} & - & Normal; RSS of L and D \\
44 & EET~83309 (U) \cite{burton2012propensity} & - & Normal; RSS of L and D \\
45 & LEW~85328 (U) \cite{burton2012propensity} & - & Normal; RSS of L and D \\
46 & GRA~95205 (U) \cite{burton2012propensity} & - & Normal; RSS of L and D \\
47 & LAR~04315 (U) \cite{burton2012propensity} & - & Normal; RSS of L and D \\
48 & ALH~84028 (CV3) \cite{burton2012propensity} & - & Normal; RSS of L and D \\
49 & EET~96026 (CV3) \cite{burton2012propensity} & - & Normal; RSS of L and D \\
50 & LAP~02206 (CV3) \cite{burton2012propensity} & - & Normal; RSS of L and D \\
51 & GRA~06101 (CV3) \cite{burton2012propensity} & - & Normal; RSS of L and D \\
52 & LAR~06317 (CV3) \cite{burton2012propensity} & - & Normal; RSS of L and D \\
53 & ALHA~77307 (CO3) \cite{burton2012propensity} & - & Normal; RSS of L and D \\
54 & MIL~05013 (CO3) \cite{burton2012propensity} & - & Normal; RSS of L and D \\
55 & DOM~08006 (CO3) \cite{burton2012propensity} & - & Normal; RSS of L and D \\
56 & Allende (CV3) \cite{burton2012propensity} & - & Normal; RSS of L and D \\
57 & Ivuna (CI) \cite{burton2014effects} & - & Normal; RSS of L and D \\
58 & SM2 (CM2) \cite{burton2014effects} & - & Normal; RSS of L and D \\
59 & SM12 (CM2) \cite{burton2014effects} & - & Normal; RSS of L and D \\
60 & SM51 (CM2) \cite{burton2014effects} & - & Normal RSS of L and D \\
61 & E.~coli (MIC) \cite{simensen2022experimental} & - & Normal; Reported absolute uncertainties \\
62 & M.~thermophila~WT (MIC) \cite{jiang2024} & - & Normal; Reported absolute uncertainties \\
63 & Geobacillus~LC300 (MIC) \cite{cordova201713c} & - & log-normal (uncertainties not reported; assumed relative error of 5\%) \\
64 & Thermus~thermophilus~HB8 (MIC) \cite{cordova201713c} & - & log-normal (uncertainties not reported; assumed relative error of 5\%) \\
65 & Rhodothermus~marinus~DSM~4252 (MIC) \cite{cordova201713c} & - & log-normal (uncertainties not reported; assumed relative error of 5\%) \\
66 & Methanococcus~maripaludis (MIC) \cite{wang2024microbial} & - & Normal; Reported absolute uncertainties \\
67 & Synechococcus~7002 (MIC) \cite{beck2018measuring} & - & Normal; Reported absolute uncertainties \\
68 & A.~acidocaldarius~DSM~446 (MIC) \cite{beck2018measuring} & - & Normal; Reported absolute uncertainties \\
69 & Kara Sea Surf. (MIX)\cite{gaye2022aaah} & 9 replicates at surface waters & $t$-dist; Averaged over 9 replicates  \\

\hline
\end{tabular}
\end{sidewaystable}

\FloatBarrier
\section{Pairwise dissimilarity tests}

\subsection*{Maximum $p$-value across diversity orders}

To evaluate the behavior of our rank-based test, we compared it against two classical nonparametric alternatives: the Mann–Whitney 
$U$ and Kolmogorov–Smirnov (KS) tests.\cite{mann1947test, massey1951kolmogorov} Each was applied pointwise across the diversity domain, and each sample pair was summarized by the maximal corrected $p$-value observed over 
$q$ (see Methods). This is the same evaluative framework used in the main analysis, differing only in the choice of test statistic. The Mann–Whitney test captures relative shifts in central tendency; the KS test responds to cumulative rank differences. Both are nonparametric and widely used, providing natural points of comparison. Results are shown in section~\ref{fig: max_mwu} and section~\ref{fig: max_ks}, respectively.

\begin{figure*}[b!]
\centering
\rotatebox[origin=c]{0}{\includegraphics[scale = 0.8]{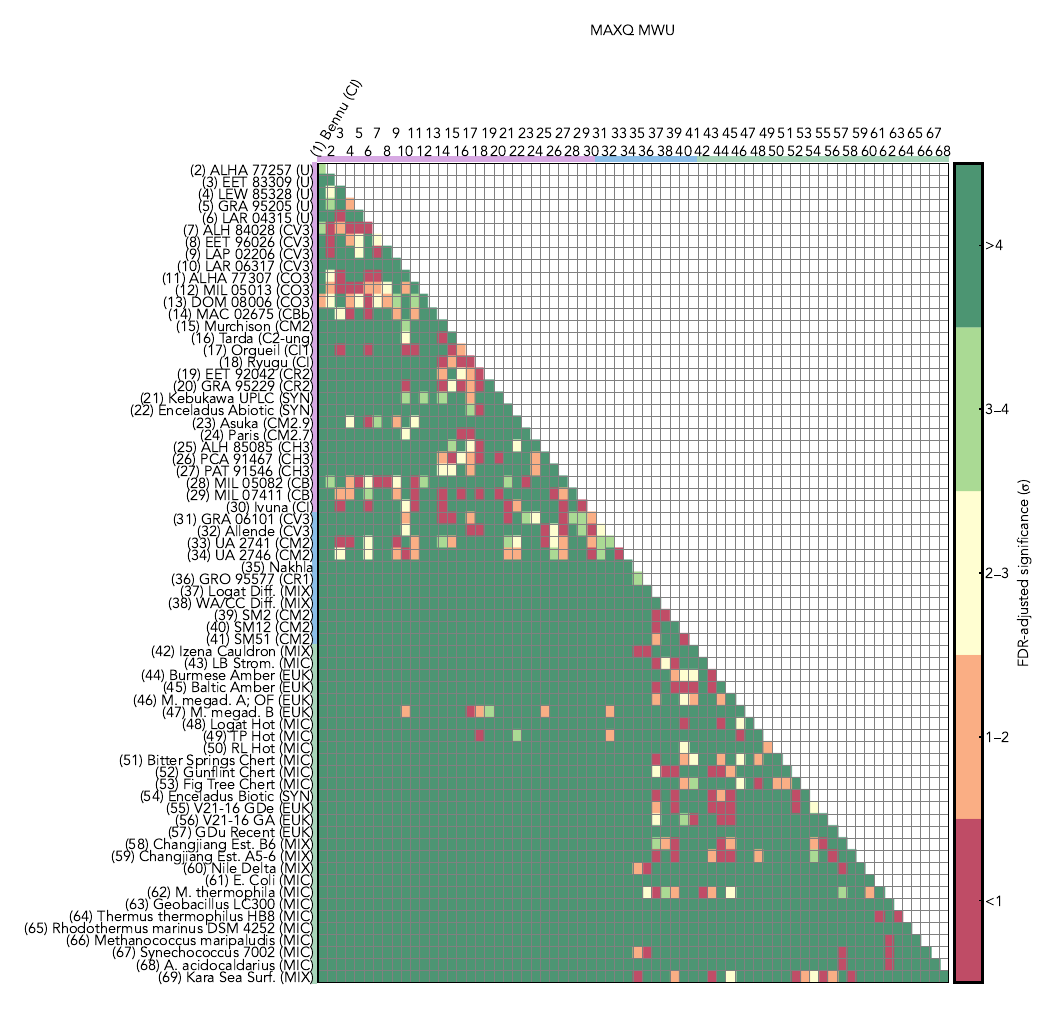}}
  \caption{\textbf{Pairwise significance matrix of evenness curves using the Mann-Whitney $U$ test.} Formatting similar to Extended Data Figs. 1--2.}
     \label{fig: max_mwu}
\end{figure*}

\begin{figure*}[b!]
\centering
\rotatebox[origin=c]{0}{\includegraphics[scale = 0.8]{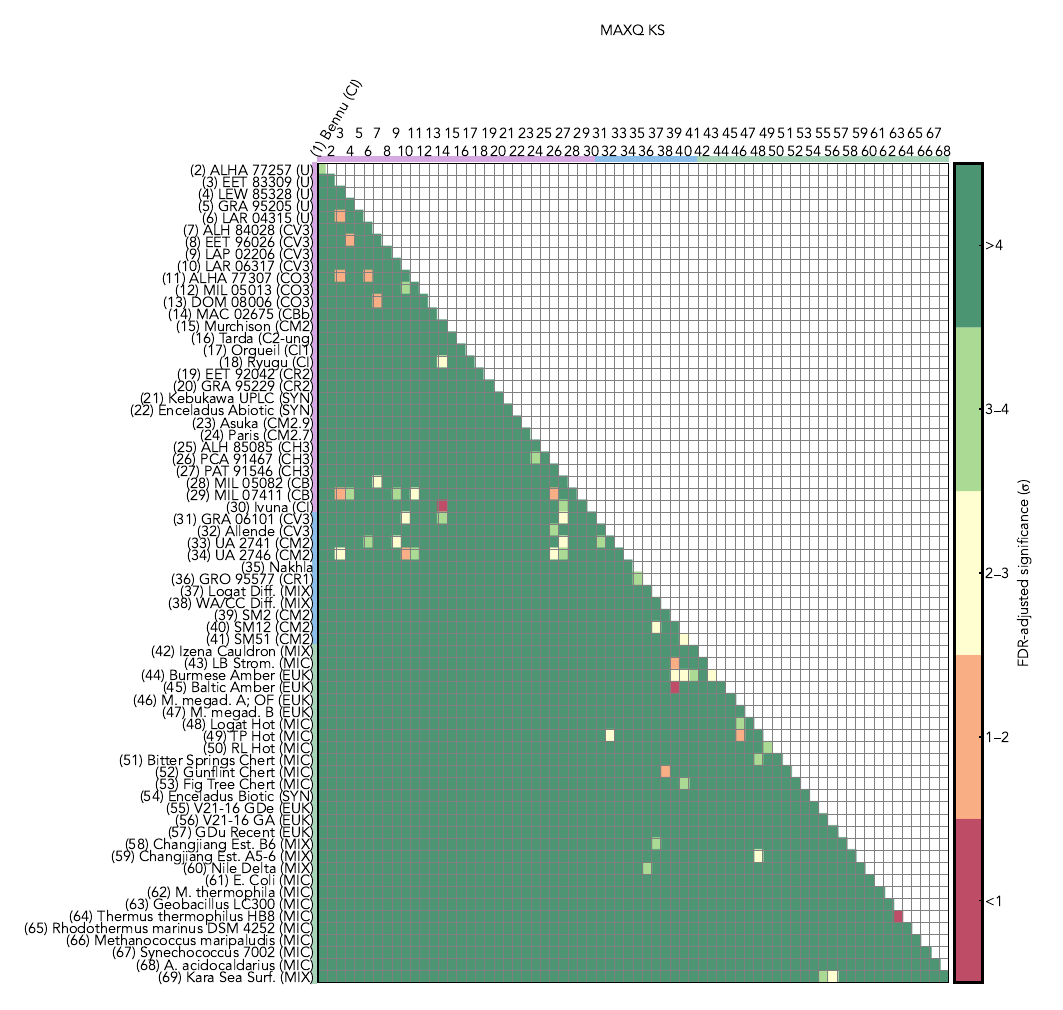}}
  \caption{\textbf{Pairwise significance matrix of evenness curves using the Kolmogorov-Smirnov test.} Formatting similar to Extended Data Figs.~1--2.}
     \label{fig: max_ks}
\end{figure*}

Both tests produce dissimilarity structures that differ notably from dissimilarities estimated with the empirical $p$-value (Fig.~2). In several cases, the resulting partitions include groupings that lack compositional or contextual coherence. While these tests also evaluate each diversity order $q$ independently, they rely on classical distributional statistics that may be strongly affected by local fluctuations, even when distributions remain largely overlapping.

\subsection*{Area Under Curve (AUC) approach}

Rather than evaluating differences pointwise across the diversity domain, we also tested whether the overall shape of each evenness curve could be captured by its area under the curve (AUC). Each realization of $E(q)$ yields a scalar AUC value, producing a distribution per sample. Pairwise comparisons were then performed on these AUC distributions using four tests: our empirical overlap method, the Mann–Whitney $U$, the Kolmogorov–Smirnov, and functional ANOVA.\cite{cuevas2004anova} This approach emphasizes total evenness magnitude, smoothing over local deviations. Results are shown below.

Among the four AUC-based tests, our empirical $p$-value yielded results closely aligned with those of the main analysis (see Supplementary Figure~\ref{fig: auc_emp}), separating biotic and abiotic samples with similar structure but reduced conservatism. The remaining tests assigned uniformly high significance to all pairwise comparisons, including those with minimal compositional contrast. This outcome likely reflects the constrained geometry of evenness curves: smooth, monotonic functions over a shared domain. Small differences in amplitude or curvature accumulate systematically in the AUC metric, leading classical tests to report significant separability between each pair of samples.

\begin{figure*}
\centering
\rotatebox[origin=c]{0}{\includegraphics[scale = 0.8]{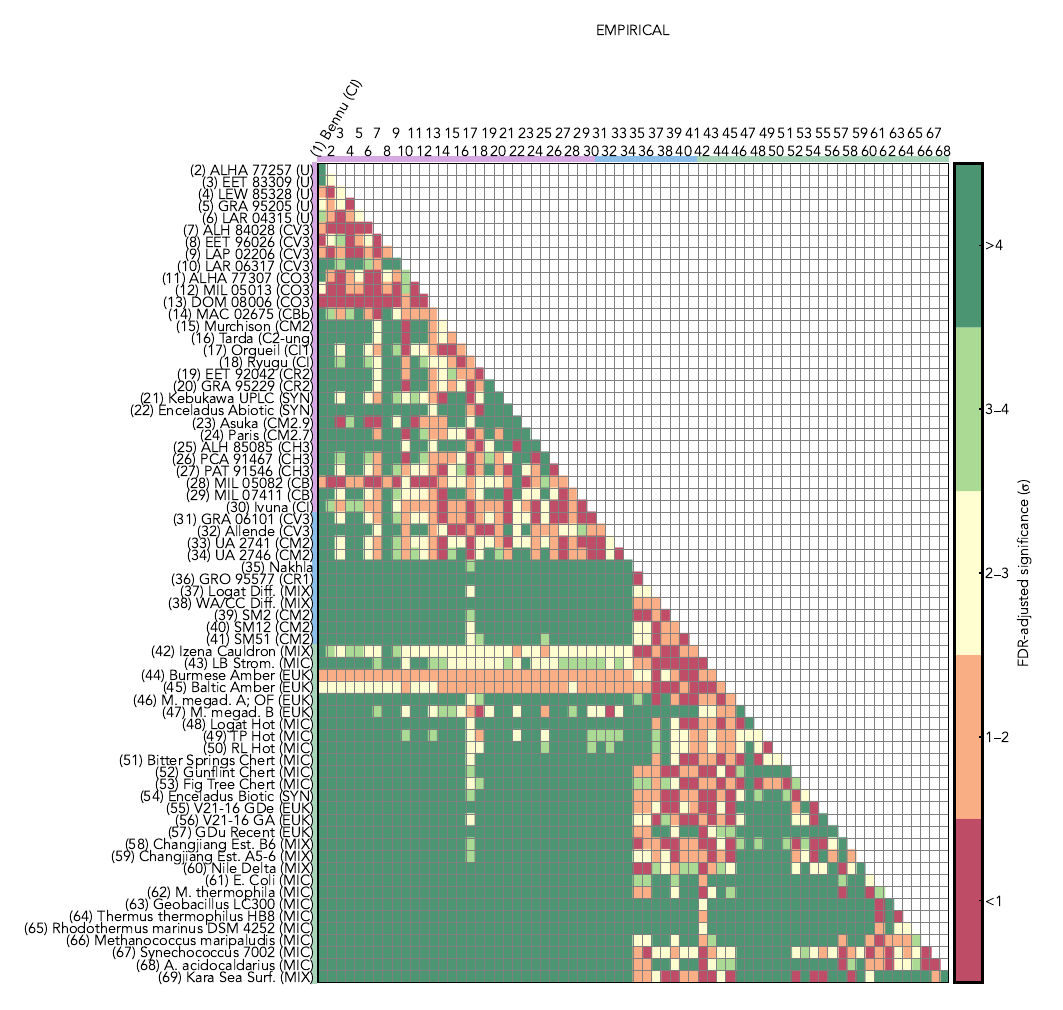}}
  \caption{Pairwise significance matrix of evenness curves, estimated with an empirical $p$-value to AUC distributions. Formatting similar to Extended Data Figs. 1--2.}
     \label{fig: auc_emp}
\end{figure*}

\begin{figure*}
\centering
\rotatebox[origin=c]{0}{\includegraphics[scale = 0.8]{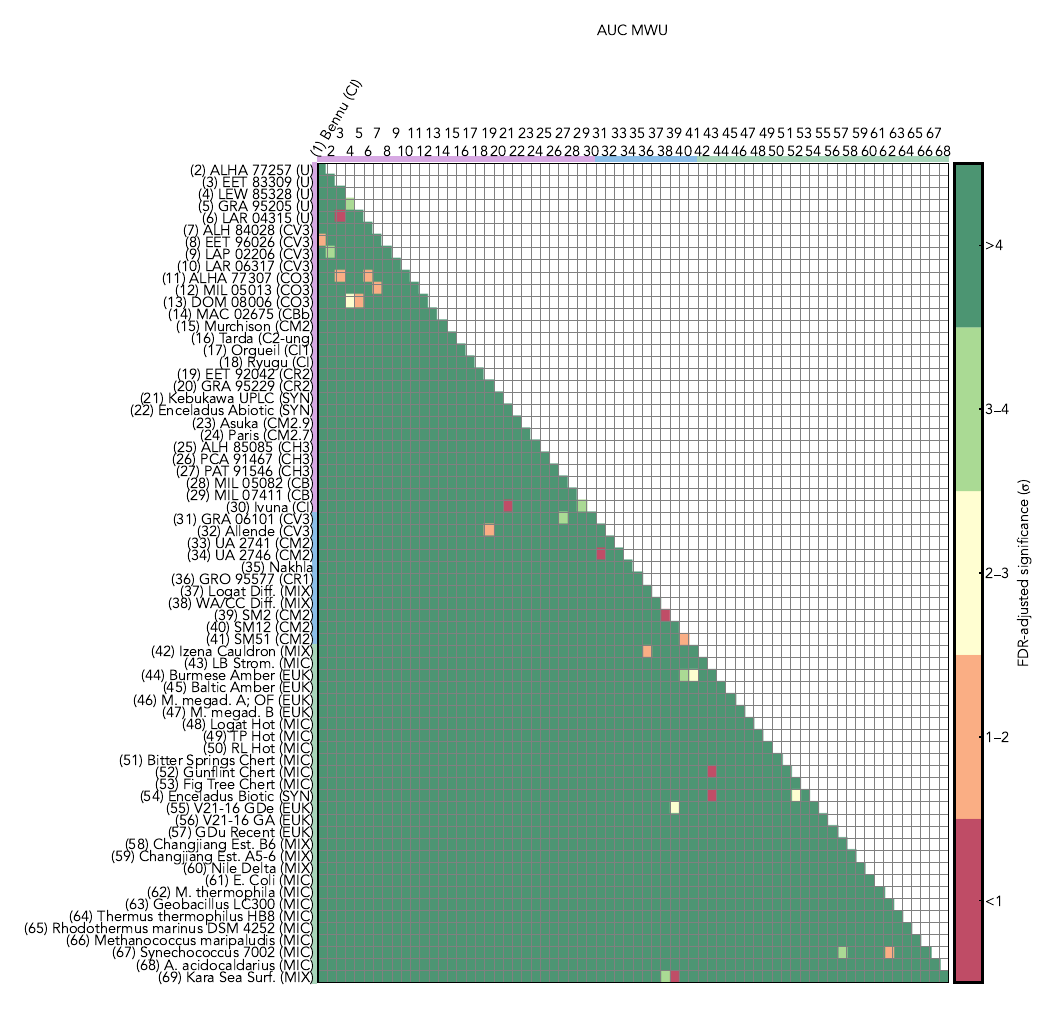}}
  \caption{Pairwise significance matrix of evenness curves, estimated through applying the Mann-Whitney test to AUC distributions. Formatting similar to Extended Data Figs. 1--2.}
     \label{fig: auc_mwu}
\end{figure*}

\begin{figure*}
\centering
\rotatebox[origin=c]{0}{\includegraphics[scale = 0.8]{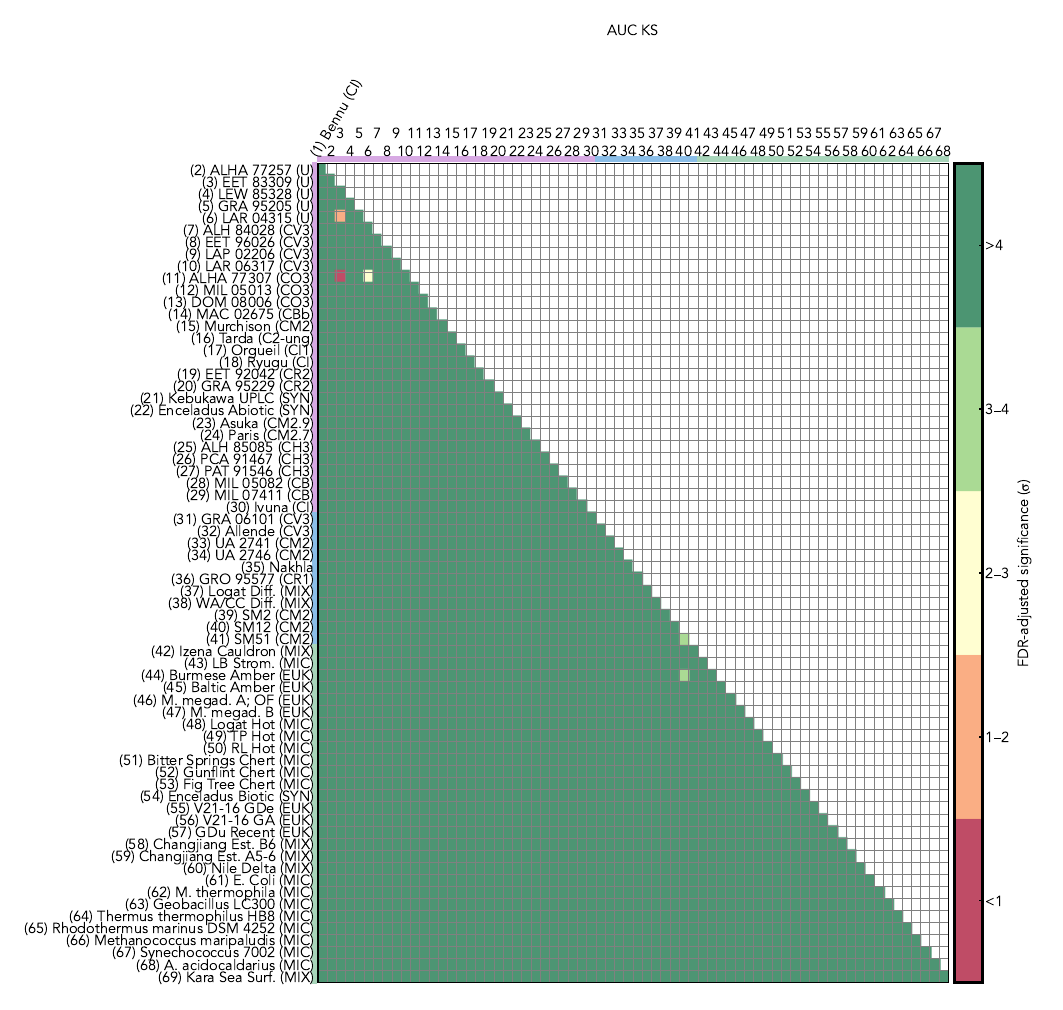}}
  \caption{Pairwise significance matrix of evenness curves, estimated through applying the Kolmogorov-Smirnov test to AUC distributions. Formatting similar to Extended Data Figs. 1--2.}
     \label{fig: auc_ks}
\end{figure*}

\begin{figure*}
\centering
\rotatebox[origin=c]{0}{\includegraphics[scale = 0.8]{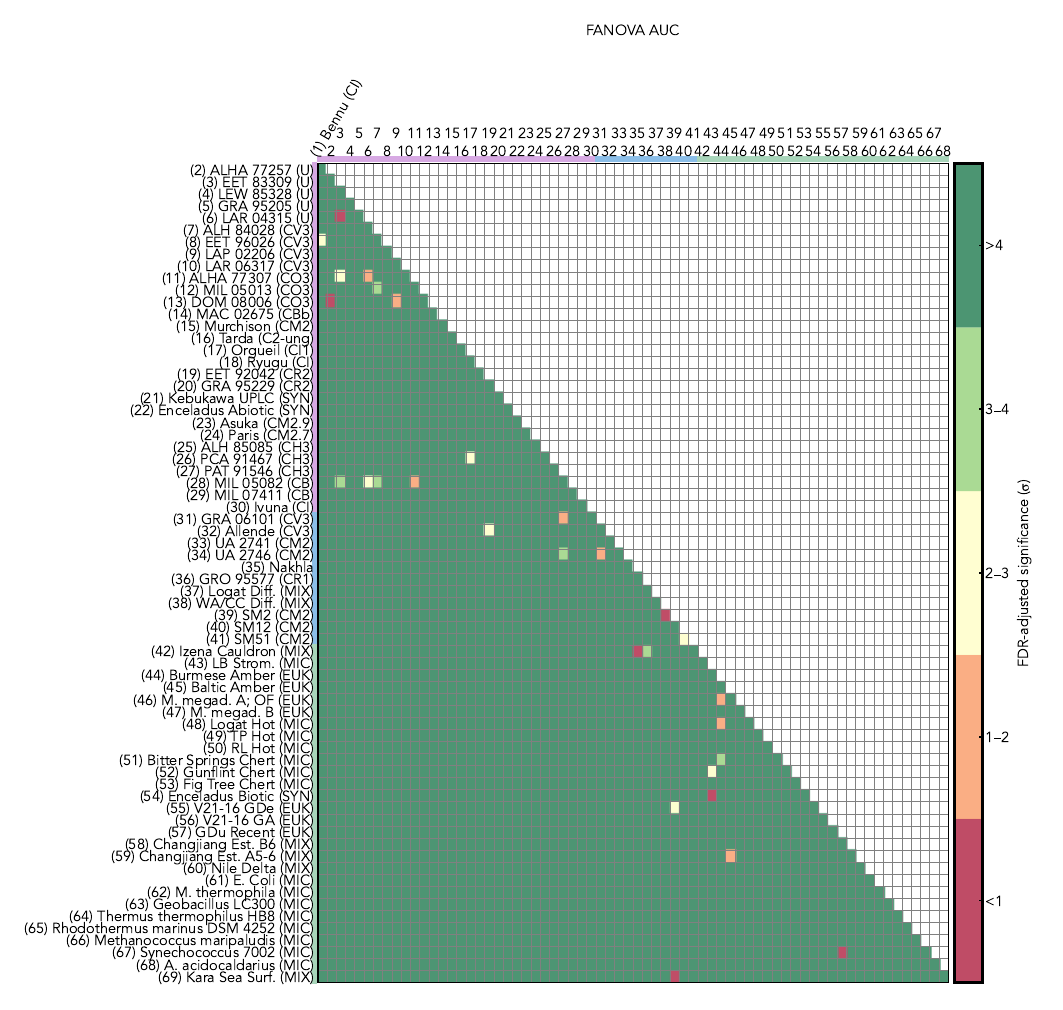}}
  \caption{Pairwise significance matrix of evenness curves, estimated through applying functional ANOVA to AUC distributions. Formatting similar to Extended Data Figs. 1--2.}
     \label{fig: auc_fanova}
\end{figure*}

\section{Diversity analysis of fatty acids} \label{app: fatty_acids}

We apply our framework to a limited dataset of 28 fatty-acid assemblies, summarized in \ref{tab: database_table_fatty}. Like amino acids, fatty acids are key building blocks of terrestrial life, but can also be produced abiotically. Fatty acids make up the lipid membranes of bacterial cells, and the abundance of individual acids varies among bacterial cultures.\cite{hildebrand1964fatty} Finding a common biotic signature in fatty-acid profiles and, thus, discriminating between biotically and abiotically produced fatty acids is critical in the search for life beyond Earth.

Reported abiotic fatty-acid datasets suitable for biosignature-style comparisons are scarce, and the overlap in chain-length coverage between biotic and abiotic studies is limited. Most biotic profiles emphasize membrane-relevant long-chain fatty acids (typically $\gtrsim$C$_{12}$), whereas abiotic measurements and models are often restricted to short-chain species (typically $\lesssim$C$_{10}$). Fatty acids also occur as both straight-chain and branched isomers, which can reflect distinct formation pathways. To avoid conflating these populations and to maintain consistent coverage across samples, we restrict our analysis to straight-chain fatty acids and to a common short-chain window: C$_{7-10}$ where available, and C$_{6-9}$ for the LON~94101 sample, in which no C$_{10}$ was detected.\citep{aponte2014chirality} This choice prioritizes comparability across abiotic and biotic datasets over breadth in chain length.

The dataset includes laboratory Fischer--Tropsch synthesis products,\citep{mccollom2007abiotic, missbach2018assessing} a modeled abiotic fatty-acid profile for the ocean of Enceladus,\citep{klenner2020analog} carbonaceous chondrites with reported monocarboxylic acid distributions,\citep{aponte2011effects, aponte2014chirality} microbial and plant lipid profiles,\citep{hildebrand1964fatty, delude2016primary, cordova201713c} and geologic lipid assemblages from soils and sediment cores spanning depths from several centimeters to nearly 3~meters.\citep{hoving2022phospholipid, gorka2023beyond, meyer2024rdbc} Since none of these sources report measurement uncertainties in a consistent manner across the full dataset, we adopt a conservative 10\% relative error and assume log-normal uncertainties for all reported abundances, with three exceptions. For the soil samples from Gorka et al. (2023),\citep{gorka2023beyond} which report four replicate measurements per sample, uncertainties are instead represented using a Student’s $t$ distribution constructed from the reported replicates. For the Fischer–Tropsch synthesis from Missbach et al. (2018),\citep{missbach2018assessing} we adopt a 20\% relative uncertainty, consistent with the experiment-to-experiment variability in relative abundances reported in their Table 1.

We analyze dissimilarities among the fatty-acid assemblages using the same framework applied to amino acids. Analysis results are shown in Fig.~3. Abiotic and biotic samples form two separate clusters and can be distinguished with high fidelity. Biotic samples exhibit lower evenness, consistent with preferential production of even-numbered fatty acids in biochemical systems, whereas abiotic samples are comparatively more uniform across chain lengths. Notably, the ALH~84033 and ALH~84034 samples are the least disambiguated from the biotic samples, consistent with a more uneven short-chain distribution (dominated by C$_7$)\citep{aponte2011effects} rather than an even-carbon bias.
Within the meteoritic subset, monocarboxylic acids partition into two structurally distinct populations: straight-chain and branched isomers. Their relative abundances vary across samples and with carbon number, and prior meteoritic analyses show that branched/straight ratios shift systematically with alteration state.\citep{aponte2011effects, alexander2017nature, lai2019meteoritic} The two subsets therefore need not reflect a single, tightly coupled production history across the full chain-length range. Treating them as a unified molecular family mixes partially decoupled abundance structures, such that differences in their relative amplitudes can project as artificial dominance in the diversity metric.
To isolate this effect, we perform the dissimilarity analysis for three representations of each meteoritic assemblage: the combined set (\ref{fig: FA_both}), straight-chain species only (\ref{fig: FA_straight}), and branched species only (\ref{fig: FA_branched}). When evaluated independently, both subsets exhibit internally coherent evenness patterns and preserve separation from biotic samples. In the combined representation, the diversity signal becomes sensitive to the branched/straight partition itself, so that shifts in total subset reduce the distinction between biotic and abiotic samples.

Reported meteoritic monocarboxylic acid abundances decrease with carbon number, and higher-C members are more frequently absent in altered samples.\citep{aponte2011effects} We therefore define the chain-length intervals operationally, based on stable representation in the compiled dataset rather than an assumed mechanistic cutoff. In our compilation, branched isomers are reproducibly represented over C$_{6-9}$, whereas branched C$_{10}$ is intermittently reported and acts as a high-leverage tail component in the diversity analysis. Straight-chain species, by contrast, remain reproducibly represented over C$_{7-10}$. This transition lies near the carbon-number regime where straight- and branched-acid abundances were reported to lose tight coupling (with correlations holding for C$\lesssim 7$), which motivates treating the two subsets separately beyond the short-chain range.\citep{lai2019meteoritic}
Including carbon numbers beyond these supported intervals introduces intermittent non-detections that bias evenness under log-normal uncertainty propagation. We therefore use branched C$_{6-9}$ and straight-chain C$_{7-10}$ as the primary subset definitions, and present the combined representation separately.

The biotic samples of a geologic context exhibit a systematic gradient correlated with burial depth. Shallow sediment and soil samples cluster closer to fresh microbial lipid profiles, consistent with active or recently active biomass inputs, whereas more deeply buried samples shift toward lower evenness and overlap with thermophilic microbial assemblages such as \textit{Thermus thermophilus} HB8 and \textit{Rhodothermus marinus} DSM~4252.\citep{cordova201713c} This trend is consistent with progressive diagenetic filtering, whereby labile lipid components are preferentially removed with depth and temperature, leaving a residual distribution increasingly dominated by saturated straight-chain fatty acids characteristic of thermophilic membranes and subsurface-adapted metabolisms.

At the same time, these results should be interpreted with appropriate caution, because published abiotic and biotic datasets rarely report fully overlapping chain-length windows or isomer classes, and our analysis therefore compares a restricted subset of straight-chain species rather than complete lipid inventories. Even so, the distribution of fatty-acid samples already exhibits structured trends, analogous to the preservation and processing gradients observed for amino acids, suggesting that biologically and environmentally meaningful information is retained in lipid assemblages.

\begin{figure*}[b!]
\centering
\rotatebox[origin=c]{0}{\includegraphics[scale = 0.5]{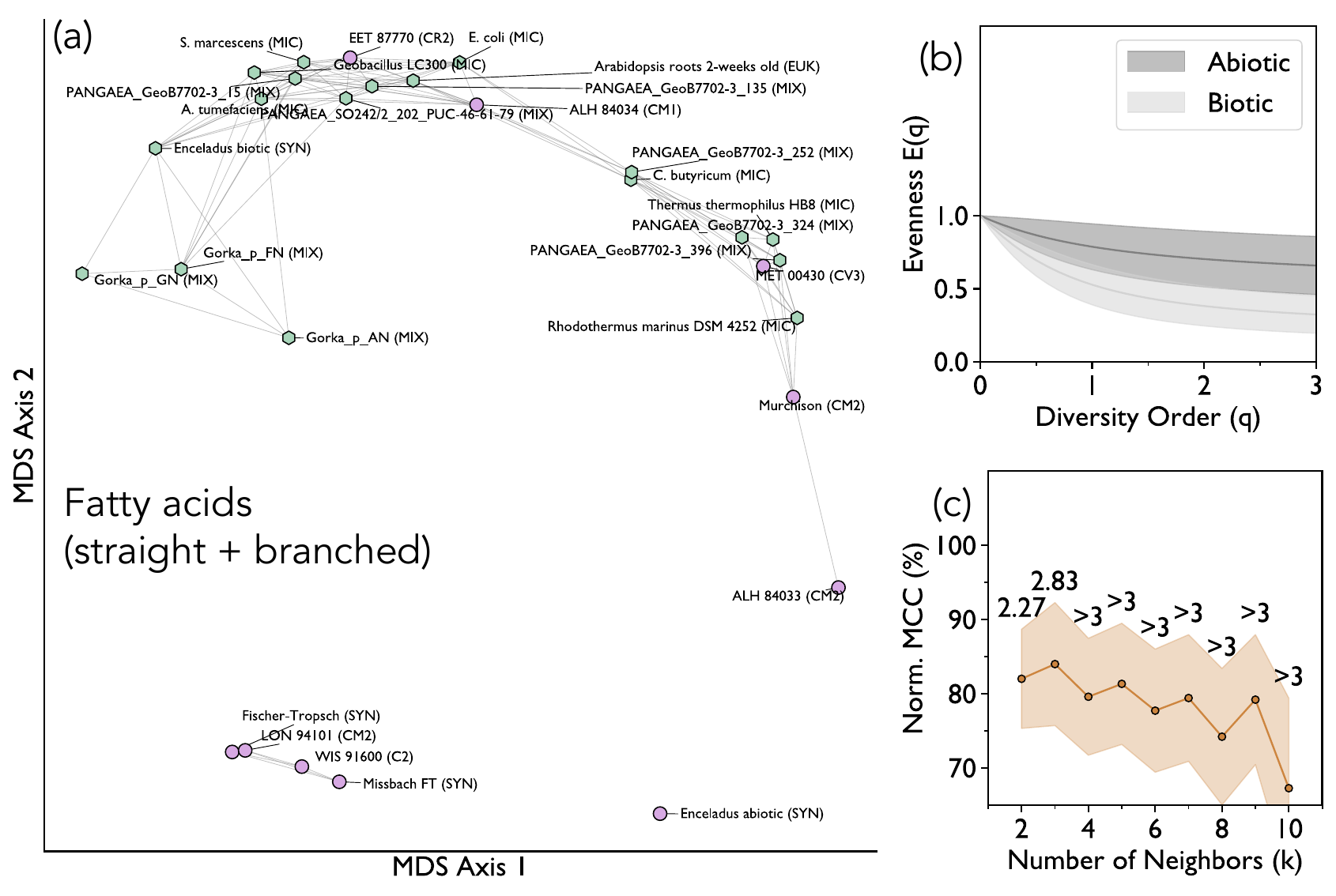}}
\caption{\textbf{Dissimilarity analysis of evenness curves of fatty-acid assemblages, similar to the amino acid analysis shown in Figs. 2 and 3. Abiotic samples from meteorites contain only straight-chain species.}}
\label{fig: FA_both}
\end{figure*}

\begin{figure*}[b!]
\centering
\rotatebox[origin=c]{0}{\includegraphics[scale = 0.5]{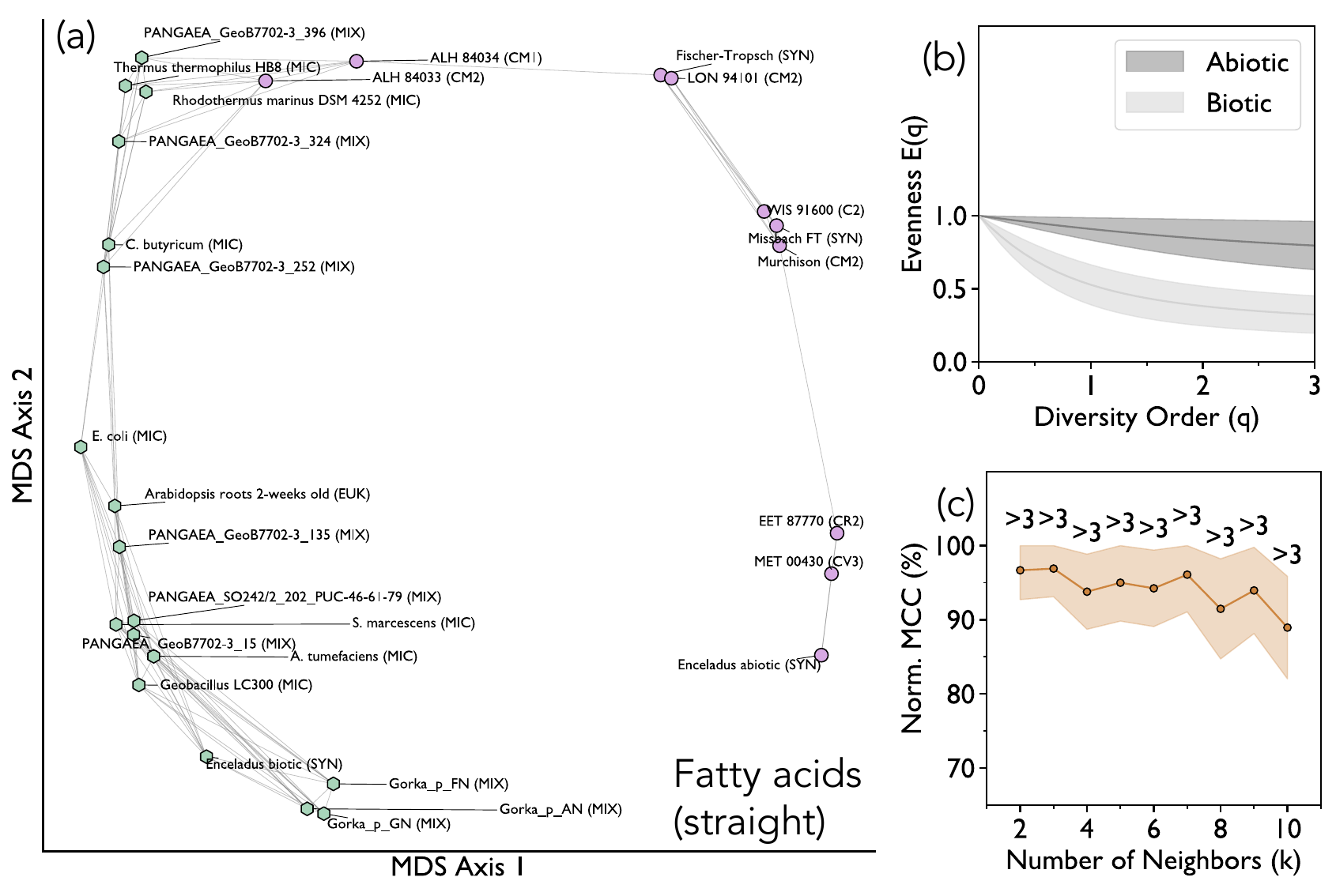}}
\caption{\textbf{Dissimilarity analysis of evenness curves of fatty-acid assemblages, similar to the amino acid analysis shown in Figs. 2 and 3. Abiotic samples from meteorites contain only straight-chain species.}}
\label{fig: FA_straight}
\end{figure*}

\begin{figure*}[b!]
\centering
\rotatebox[origin=c]{0}{\includegraphics[scale = 0.5]{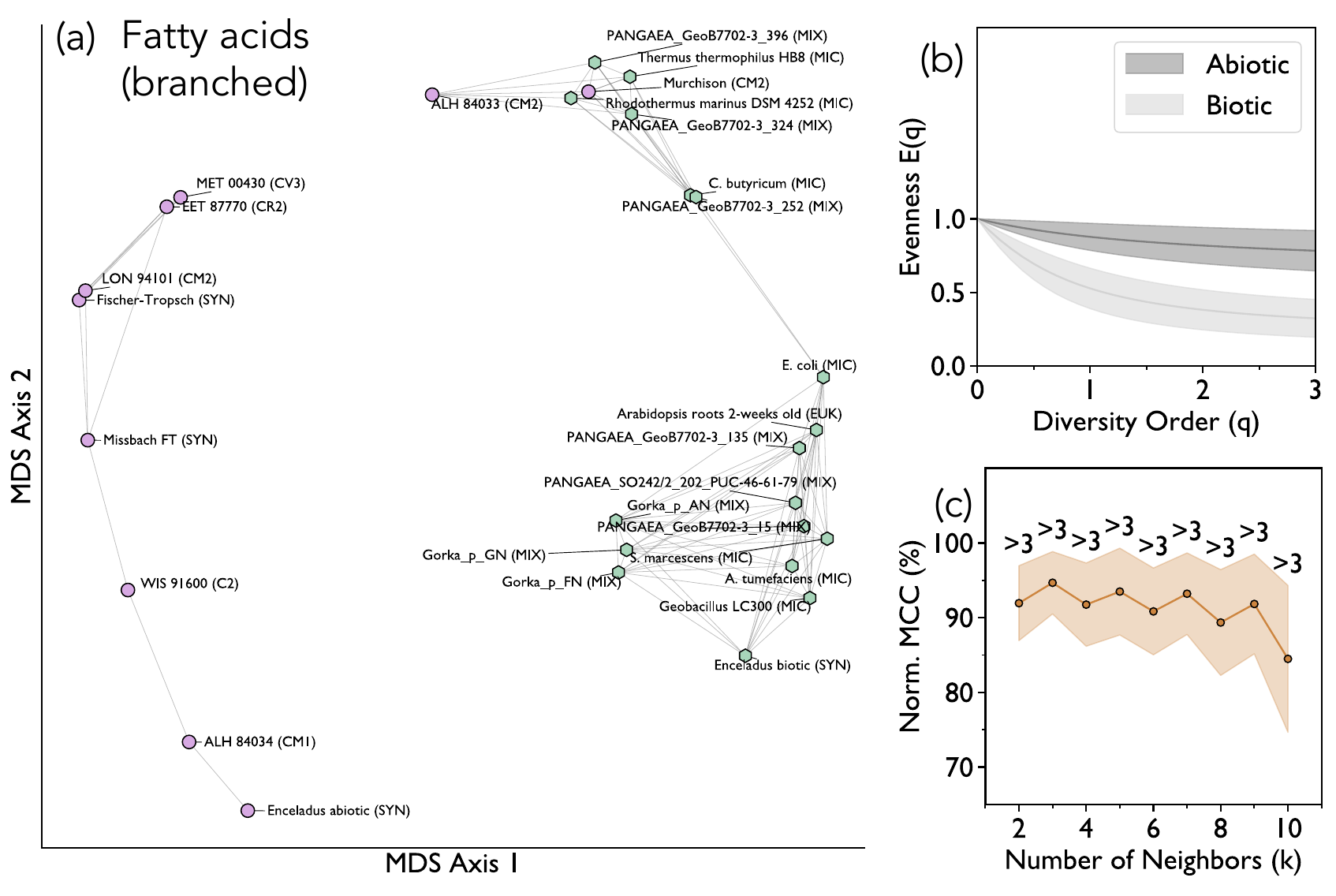}}
\caption{\textbf{Dissimilarity analysis of evenness curves of fatty-acid assemblages, similar to the amino acid analysis shown in Figs. 2 and 3. Abiotic samples from meteorites contain only straight-chain species.}}
\label{fig: FA_branched}
\end{figure*}

\section{Bootstrapping and hypothesis-testing $k$NN}
\label{app:bootstrap_knn}

To quantify the robustness of compositional separability under sampling variability, we evaluate classification performance using a bootstrap-based $k$-nearest-neighbor ($k$NN) classification. The analysis is designed to propagate finite-sample uncertainty while testing whether observed classification skill exceeds that expected under random labeling.
We begin from a fixed pairwise dissimilarity matrix derived from the evenness-curve representation. For each bootstrap realization, we draw a class-balanced subsample of biotic and abiotic profiles without replacement. Classification is performed in the resulting subspace using a $k$NN classifier with $k$ neighbors, and performance is quantified using the Matthews correlation coefficient (MCC), which remains well-defined under class imbalance.
This procedure is repeated across 100 bootstrap realizations, yielding a distribution of MCC values for each $k$. The mean MCC summarizes expected classification performance under resampling, while the spread reflects sensitivity to finite-sample fluctuations.

To assess the statistical significance of the mean MCC for each $k$, we construct a null distribution by repeating the same bootstrap procedure after randomly permuting the class labels assigned to samples. For each $k$, the observed mean MCC is compared to the permutation-based null, consisting of 20,000 permutation rounds, and reported as a $z$-score measuring the deviation, in units of the null standard deviation, from label-randomized expectations.
Together, this framework distinguishes between sampling variability and statistical significance. Bootstrap dispersion quantifies uncertainty in achievable classification performance, while permutation-based $z$-scores test whether separability reflects genuine compositional structure rather than chance alignment.

\section{Modeling radiolytic degradation on Europa} \label{app: europa_model}

We assess the survivability of the diversity signal under radiolysis on Europa, a mechanism shown to be the dominant degradation mechanism affecting amino acids in near-surface ice.\cite{yoffe2025fluorescent}
We adopt a forward model that links cumulative radiation dose to time-dependent compositional change. Each amino acid species is assumed to degrade independently via first-order exponential decay. The abundance of the $i^{\rm{th}}$ species at time~$t$ follows
\begin{equation} \label{eq:radiolysis_decay}
    A_i(t) = A_i(0)\, \exp(-r_i D t),
\end{equation}
where $A_i(0)$ is the initial abundance, $D$ is the local dose rate, and $r_i$ is the species-specific radiolytic constant. The dose rate, $D$, is determined by local depth and latitude and reflects energy deposition from Jovian magnetospheric particles.\cite{nordheim2018preservation, yoffe2025fluorescent, nordheim2022magnetospheric} The radiolytic constants $r_i$ vary across amino acids due to molecular structure.\cite{pavlov2024radiolytic} This variation induces selective degradation of amino acid species.

To test whether selective radiolysis can erase the diversity signal, we compare degraded biotic profiles to three baselines: their pristine state, a pristine abiotic reference, and an abiotic profile undergoing identical degradation. This construction separates generic attenuation from a collapse of biotic--abiotic separability, and tracks signal persistence continuously in time.
Radiolysis is parameterized using experimentally measured decay constants for amino acids embedded in frozen biological material. We include all 15 amino acids for which radiolytic constants and uncertainties have been reported for dead \textit{E.~coli} under Europa- and Enceladus-relevant conditions.\citep{pavlov2024radiolytic} Each amino acid evolves independently under equation~(\ref{eq:radiolysis_decay}), with decay constants sampled from their reported uncertainty ranges (see the Supplementary Material of Pavlov et al.~2024\citep{pavlov2024radiolytic}). These constants are applied to both biotic and abiotic profiles, since no comparable radiolysis constants are available for abiotic amino acids across this species set. Differences in evolution, therefore, arise solely from differences in initial relative abundances.

\subsection*{Radiolysis model} \label{app_europa_experimental_setup}

We consider all biotic and abiotic samples in the dataset compiled during this study that contain at least 7 amino acids and have available radiolytic constants. Samples labeled as \textit{Biotic Degraded} are excluded to avoid conflating radiolytic evolution with prior diagenetic modification. This selection yields 16 biotic and 12 abiotic samples.
To quantify statistical persistence, we perform 100 simulations. In each simulation, we randomly draw one eligible biotic and one eligible abiotic sample, and propagate 10,000 uncertainty-derived abundance realizations for each sample under radiolytic degradation. Evolution is evaluated at a nominal leading-hemisphere location (60$^\circ$ latitude, 90$^\circ$W longitude) at three near-surface depths: 1, 10, and 50~millimeters. Radiolytic constants are drawn once per simulation from their reported uncertainties and applied to all realizations; profiles are renormalized to unit total abundance and mapped to evenness curves at each time step.
At each time, the degraded biotic evenness-curve distribution is compared to three benchmarks: (\textbf{1}) the pristine biotic distribution, (\textbf{2}) the pristine abiotic distribution, and (\textbf{3}) the simultaneously degrading abiotic distribution. Each comparison is summarized as a dissimilarity $z$-score, yielding depth-dependent time series that quantify loss of biotic structure, drift toward abiotic structure, and residual biotic--abiotic separability under matched degradation.

\subsection*{Energy deposition by magnetospheric particles} \label{app_ice_Edep}

To model the radiation environment at Europa’s surface, we adopt the energy deposition profiles presented in Yoffe et al. (2025),\cite{yoffe2025fluorescent} who simulated charged particle transport through near-surface ice using \texttt{G4beamline},\cite{roberts2007g4beamline} a Monte Carlo particle physics toolkit. These simulations account for the depth-dependent energy flux from electrons and the three dominant magnetospheric ions (p, O$^{2+}$, and S$^{3+}$), incorporating particle power spectra derived from the measurements by the \textit{Voyager} and \textit{Galileo} missions, modulated by magnetospheric drift patterns specific to Europa’s leading hemisphere.\cite{paranicas2001electron, nordheim2018preservation, nordheim2022magnetospheric} The resulting energy deposition rates are depth- (down to one meter) and location-resolved.
For a detailed presentation of the simulations and corresponding results, see Appendix A of Yoffe et al. (2025).\citep{yoffe2025fluorescent} The depth- and location-resolved dose rate maps are available online in Zenodo.\citep{yoffe2026_molecular_diversity_zenodo}


\end{supplementary}

\end{document}